\documentclass{raa}            

\usepackage{graphicx,times,color,amsmath,amssymb}             
\input{epsf.sty}                        
\input{psfig.sty}                       

\begin{document}

   \title{A magnetic reconnection model for quasi-periodic oscillations in black hole systems
}

   \volnopage{Vol.0 (200x) No.0, 000--000}      
   \setcounter{page}{1}          

   \author{Chang-Yin  Huang
      \inst{1,2}
   \and Ding-Xiong  Wang
      \inst{1}$^{*}$
    \and Jiu-Zhou  Wang
      \inst{1}
    \and Zhi-Yun Wang
      \inst{3}
   }

   \institute{School of Physics, Huazhong University of Science and Technology, Wuhan 430074, China; \\
        \and
            School of Mathematics and Statistics, Huazhong University of Science and Technology, Wuhan 430074, China; \\
        \and
             School of Physics and Electronic Engineering, Hubei University of Arts and Science, Xiangyang 441053, China\\
             {$^{*}$dxwang@mail.hust.edu.cn}\\
}

   \date{Received 2012 month day; accepted 2012 month day}

\abstract{ The quasi-periodic oscillations (QPOs) in black hole (BH) systems of different scales are interpreted based on the magnetic reconnection of the large-scale magnetic fields generated by the toroidal electric currents flowing in the inner region of accretion disk, where the current density is assumed to be proportional to the mass density of the accreting plasma. The magnetic connection (MC) is taken into account in resolving the dynamic equations of the accretion disk, in which the MC between the inner and outer disk regions, the MC between the plunging region and the disk, and the MC between the BH horizon and the disk are involved. It turns out that the single QPO frequency of several BH systems of different scales can be fitted by invoking the magnetic reconnection due to the MC between the inner and outer regions of the disk, where the BH binaries XTE J1859+226, XTE J1650-500 and GRS 1915+105 and the massive BHs in NGC 5408 X-1 and RE J1034+396 are included. In addition, the X-ray spectra corresponding to the QPOs are fitted for these sources based on the typical disk-corona model.
 }

   \authorrunning{C. Y. Huang, D. X. Wang, J. Z. Wang \& Z. Y. Wang}            
   \titlerunning{A magnetic reconnection model for quasi-periodic oscillations in black hole systems}  

  \maketitle

%
%
\section{Introduction}           
\label{sect:intro}

As is well known, X-ray quasi-periodic oscillation (QPO) is a common phenomenon in the radiation from black-hole binaries (BHBs). The high-frequency QPOs (HFQPOs) have been observed in several BHBs, some of which show interesting 3:2 frequency pairs (GRS 1915+105, GRO J1655-40, XTE J1550-564, H1743-322), and others display single QPO (XTE J1859+226, XTE J1650-500). QPOs have also been observed in ultraluminous X-ray sources (ULXs), e.g., a 54 mHz QPO in M82 X-1 and a 20 mHz QPO in NGC 5408 X-1 were discovered respectively by Strohmayer \& Mushotsky (\cite{strohmayer03}) and Strohmayer et al. (\cite{strohmayer07}) with XMM-Newton. Strohmayer \& Mushotsky (\cite{strohmayer09}) discovered another strong 10 mHz QPO in NGC 5408 X-1, and found that the correlation of timing and spectral properties of this source is similar to those of Galactic BHBs.

   The first convincing QPO of active galactic nuclei (AGNs) was discovered by Gierlinski et al. (\cite{gierlinski}) in narrow line Seyfert 1 RE J1034+396, which opened a window for the comparative timing studies of stellar-mass and supermassive black holes (BHs). The ~1-hour X-ray QPO observed in RE J1034+396 is analogous to the 67 Hz QPO in BHB GRS 1915+105 (Middelton \& Done \cite{middleton10}). In earlier years, quasi-periodic signals like QPOs were discovered in some supermassive BHs. For example, the power density spectra of two X-ray flares of Sgr A* observed in 2000 and 2002 show five distinct peaks at periods of ~100, 219, 700, 1150, 2250 seconds (Aschenbach et al. \cite{aschenbach}) and a quasi-periodic flux modulation with a period of 22.2 minutes was discovered in the X-ray data of the Sgr A* flare in 2004 (B¨¦langer et al. \cite{belanger}).

   A number of models have been proposed to interpret HFQPOs in BHBs. However, none of them can fully explain the characteristics of QPOs, especially the correlations of spectral and timing properties (Remillard \& McClintock \cite{remillard}; Maitra \& Miller \cite{maitra}). As is well known, HFQPOs in BHBs are strongly correlated to steep power law (SPL) state. A successful model of QPOs should link the corresponding spectral state, allowing for a highly dynamical interplay between thermal and nonthermal processes with the mechanisms operating over a wide range of luminosity (Remillard \& McClintock \cite{remillard}). QPOs observed in massive BHs are probably related to the bright state which is similar to the SPL state of BHBs (Strohmayer \& Mushotsky \cite{strohmayer03}, \cite{strohmayer09}; Middelton et al. \cite{middleton09}), and they may have the same origin as the HFQPOs in BHBs (Gierlinski et al. \cite{gierlinski}; Bian \& Huang \cite{bian}).

   Inverse Compton scattering is generally thought to be the promising radiation mechanism of SPL state (Zdziarski \cite{zdziarski}; McClintock \& Remillard \cite{mr06}, hereafter MR06), and disk-corona model is successful in fitting the spectrum of SPL state (Gan et al. \cite{gan09}, hereafter G09; Huang et al. \cite{huang}). Zhao et al. (\cite{zhao}) (hereafter Z09) interpreted HFQPOs in BHBs as the magnetic reconnection of large-scale magnetic fields generated by toroidal electric currents flowing in the disk without considering the spectral state. In this paper, we improve the model of Z09 based on the disk-corona model with the magnetic connection (MC) effects. We fit both the QPO frequencies and the corresponding X-ray spectra of four BH systems of different scales, in which two BHBs (XTE J1859+226, XTE J1650-500), one ULX (NGC 5408 X-1) and one supermassive BH (RE J1034+396) are included.

   This paper is organized as follows. A description of the model is given in Section 2. The QPO frequencies and X-ray spectra of the four sources are fitted based on the disk-corona model with MC effects in Section 3. And some issues related to our model are discussed in Section 4. Throughout this paper, the geometric units $G = c = 1$ are used.


\section{MODEL DESCRIPTION}
\label{sect:Obs}
\subsection{Origin of magnetic fields in BH systems}

Large-scale magnetic fields play very important roles in high energy astrophysical phenomena. The relativistic jets from AGNs and BHBs are launched and collimated by invoking the open large-scale magnetic fields related to the BZ and BP processes (Blandford \& Znajek \cite{bz77}; Blandford \& Payne \cite{bp82}), and the broad iron lines observed in BH systems could be interpreted by transferring energy and angular momentum from a spinning BH to its surrounding accretion disk by invoking the magnetic connection via the closed large-scale magnetic fields connecting the horizon with the inner disk (Wilms et al. \cite{wilms}; Miller et al. \cite{miller02}; Li \cite{li02a}; Wang et al. \cite{wang02}, hereafter W02). However, a consensus on the origin of large-scale magnetic fields in BH systems has not been reached.

   Unlike neutron stars, magnetic fields cannot exist on the horizon of an isolated BH, and they should be maintained by the surrounding environment, such as an accretion disk. For BHBs, the magnetic fields probably come from the plasma of the companion. One of the most promising origins of large-scale magnetic fields in BH systems is accretion disk around the BHs, and some authors calculated the magnetic field configurations by assuming the toroidal electric current in the disk (Li \cite{li02b}; Z09). However, the origin of the currents remains to be clarified.

   In this paper, we intend to interpret the origin of the electric current based on the assumption that the accreting plasma deviates somehow from electric neutrality as it flows through the Lagrange point into the Roche lobe of the BHBs, resulting in toroidal electric currents flowing in the accretion disk. In addition, we resolve the dynamical equations of the accretion disk by taking the MC effects into account and figure out the mass density and current density in the disk, and then determine the configuration of the large-scale magnetic fields by considering the interaction between the electric current and the disk with the iterative algorithm. It turns out that we can fit the association of QPOs with spectral states in BH systems of different scales based on our model.

   \begin{figure}
     \begin{minipage}[t]{0.5\linewidth}
       \centering
       \includegraphics[width=6.cm]{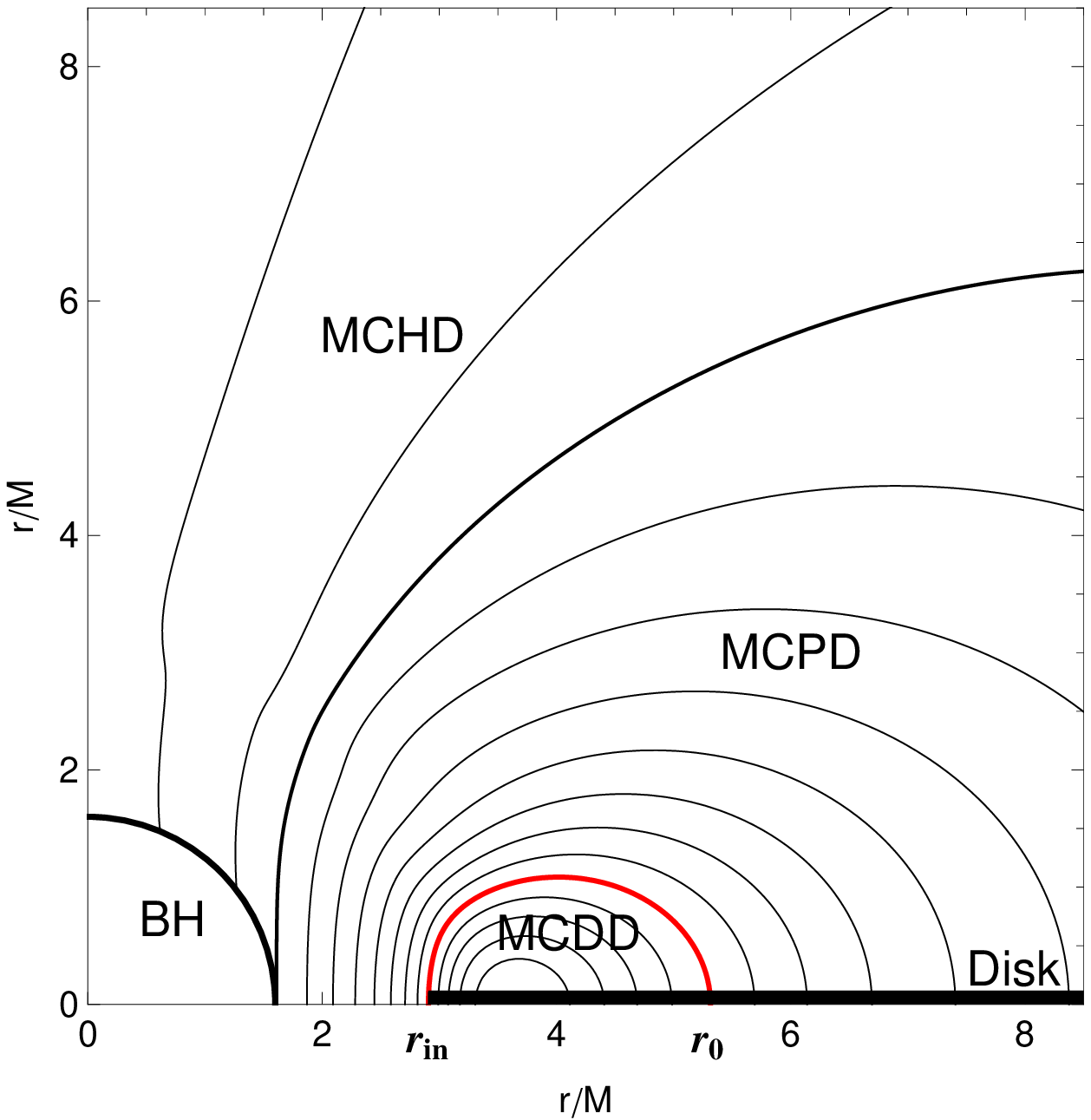}
       \caption{The configuration of large-scale magnetic fields generated by the continuous distribution of toroidal electric current flowing in the disk. The magnetic field lines are plotted in thin lines and the boundaries between three types of magnetic connection (MCHD, MCPD, and MCDD) are shown by the thick lines. The red line represents the magnetic field line connecting the radii $r_{\textup{\scriptsize in}}$ and $r_{0}$. The figure is plotted with $m_{\textup{\scriptsize BH}}=10$, $a_*=0.8$, $\dot{m}=0.1$, $\alpha=0.3$, $\eta=10^{-13}$, and $n=5$, where $\dot{m}$ and $\alpha$ are the mass accretion rate in unit of the Eddington accretion rate and the viscosity parameter, respectively.}
     \end{minipage}
     \begin{minipage}[t]{0.5\linewidth}
       \centering
       \includegraphics[width=6.cm]{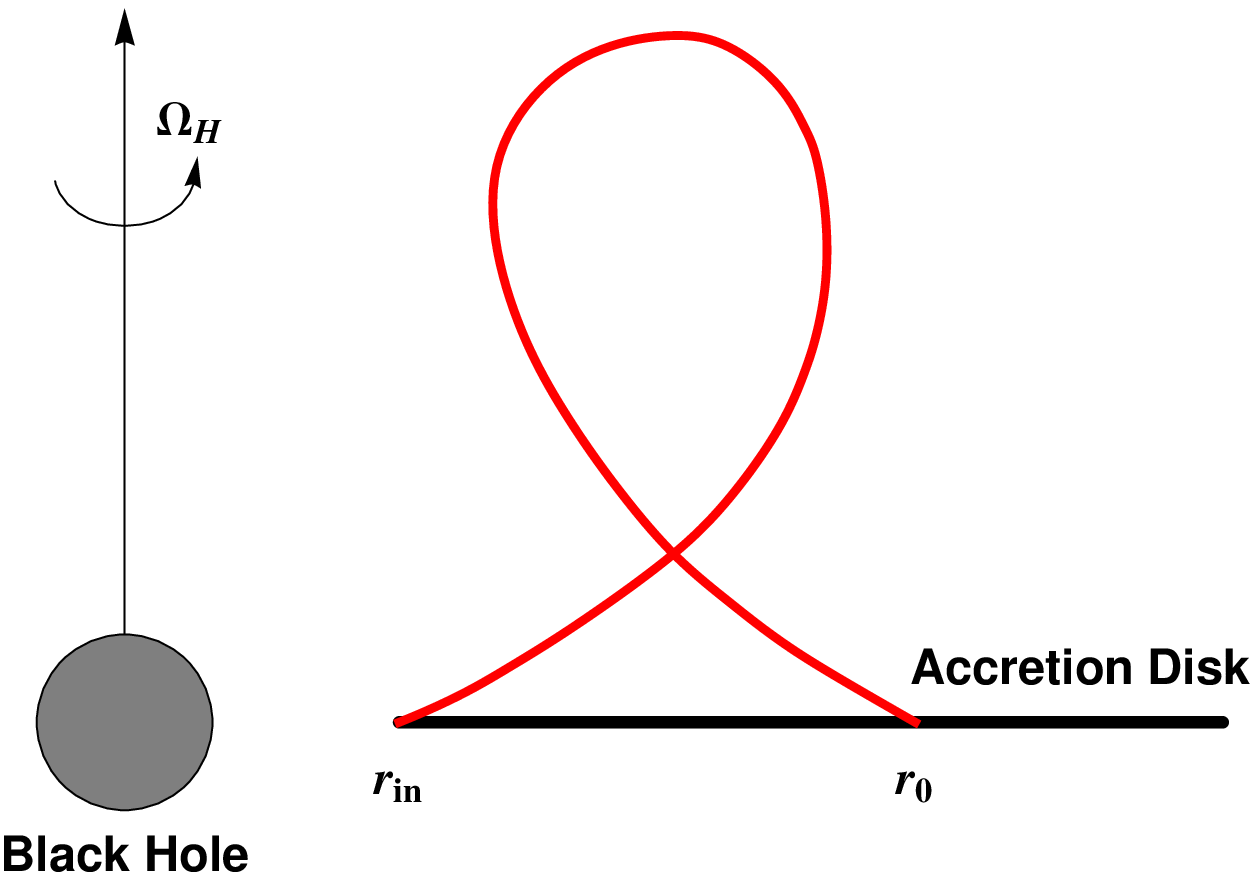}
       \caption{A schematic drawing of magnetic reconnection in MCDD, where the red line represents the field line connecting radii $r_{\textup{\scriptsize in}}$ and $r_{0}$.}
     \end{minipage}
   \end{figure}

The deviation from electric neutrality is described by defining a parameter $\eta$ as follows,
   \begin{equation}\label{eq1}
    \eta\equiv |n_{\textup{\scriptsize e}}-n_{\textup{\scriptsize p}}|/n_{\textup{\scriptsize p}},
   \end{equation}
   where $n_{\textup{\scriptsize e}}$ and $n_{\textup{\scriptsize p}}$ are the number densities of electrons and protons, respectively. The charge density can be expressed as
   \begin{equation}\label{eq2}
    \rho_{\textup{\scriptsize e}}=\eta e n_{\textup{\scriptsize p}},
   \end{equation}
   where $e=4.8\times 10^{-10}\textup{e.s.u}$ is the electron charge.

The large-scale magnetic fields generated by the toroidal electric current may in turn affect the current. For example, the field lines may pipe hot electrons into the corona above the disk therefore change the charge density in the disk. This effect should be strongest in the inner disk because the magnetic field intensity is strongest near the inner edge of the disk and decreases rapidly outwards. For simplicity, we assume this effect decreases with the increasing disk radius as a power law and the surface density of the current in the disk can be expressed as
   \begin{equation}\label{eq3}
    j=\rho_{\textup{\scriptsize e}}\cdot 2h \cdot \nu_{\textup{\scriptsize K}} \cdot r^{-n}=\eta e n_{\textup{\scriptsize p}} \cdot 2h \cdot \Omega_{\textup{\scriptsize K}}r^{1-n}=\frac{\eta e}{\mu m_{\textup{\scriptsize p}}}\rho_{\textup{\scriptsize m}}h\Omega_{\textup{\scriptsize K}}r^{1-n},
   \end{equation}
where $h$, $\rho_{\textup{\scriptsize m}}$, and $m_{\textup{\scriptsize p}}$ are the half height of the disk, the mass density and the proton mass respectively. The power law index $n$ is a free parameter for fitting the QPO frequency, and   $\mu=0.615$ is the weight-average molecular weight of the gas.
The toroidal electric current is assumed to distribute from the inner edge of the disk to the outer boundary of the disk-corona system, the radius of which is set at $r_{\textup{\scriptsize out}}=100M$ in calculations, since the radiation of the accretion disk mainly comes from the inner region.

\begin{figure}
   \centering
    {\includegraphics[width=6.5cm]{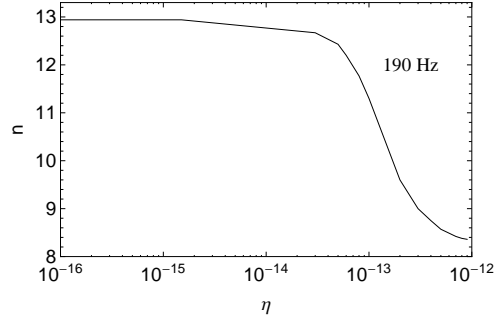}}
    \caption{The contour of 190 Hz QPO of XTE J1859+226 in $\eta-n$ parameter space with $m_{\textup{\scriptsize BH}}=12$, $a_*=0.96$, $\dot{m}=0.38$, and $\alpha=0.3$.}\label{3}
   \end{figure}

Following the work of Znajek (\cite{znajek}), Linet (\cite{linet}), Li (\cite{li02b}) and Z09, we can calculate the toroidal component of the electric vector potential determined by the current given by Eq. \eqref{3} with a given mass density of the disk matter. And the boundaries of inner-outer regions related to the magnetic field configuration can be determined. As shown by Fig.1, there are three types of magnetic field configuration generated---MC of the BH with the disk (MCHD), MC of the plunging region with the disk (MCPD), and MC of the inner and outer disk regions (MCDD). In the MCDD region, the magnetic field lines are frozen in the disk at the inner and outer footpoints. The field lines will twist themselves since the inner and outer footpoints have different angular velocities. In Fig.1, the inner and outer footpoints of the field line in red locate at $r_{\textup{\scriptsize in}}$ (the inner edge of the disk) and $r_{\textup{\scriptsize 0}}$, respectively. The value of the angular velocity difference between the footpoints of this field line is maximal among all the lines in MCDD region, so this line will twist itself first and trigger the magnetic reconnection, as shown in Fig.2, which releases magnetic energy periodically to generate flares with the frequency of the difference between Keplerian frequencies of the inner and outer footpoints. We interpret this frequency as the QPO frequency, and it reads:
   \begin{equation}\label{eq4}
   \nu_{\textup{\scriptsize QPO}} = \frac{\Omega_{\textup{\scriptsize K}}(r_{\textup{\scriptsize in}})-\Omega_{\textup{\scriptsize K}}(r_{\textup{\scriptsize 0}})}{2 \pi}
   = \nu_{\textup{\scriptsize 0}}\left[(r_{\textup{\scriptsize in}}^{3/2}M^{-3/2}+a_*)^{-1}-(r_{\textup{\scriptsize 0}}^{3/2}M^{-3/2}+a_*)^{-1}\right],
   \end{equation}
   where $\nu_0\equiv m_{\textup{\scriptsize BH}}^{-1}3.23\times10^{-4}\textup{Hz}$ and $m_{\textup{\scriptsize BH}}\equiv M/M_\odot$ is the black hole mass in unit of the solar mass. $a_*\equiv a/M$ is the dimensionless BH spin. $r_{\textup{\scriptsize in}}$ is initialized at the innermost stable circular orbit (ISCO).

   As shown by Eq. \eqref{eq3}, parameters $n$ and $\eta$ determine the current density and therefore magnetic configuration and QPO frequency once the surface density of disk matter is given. The value of $\eta$ determines the current intensity and therefore magnetic field strength, while that of $n$ determines the concentricity of the currents and magnetic fields. A larger $n$ leads to a smaller MCDD region, a smaller distance btween $r_{\rm in}$ and $r_{\rm 0}$, and hence a lower QPO frequency, while a smaller $n$ leads to the opposite results. As an example, a contour of 190 Hz QPO of XTE J1859+226 is plotted in $\eta-n$ parameter space as shown in Fig. 3. It is found that a maximum $\eta\sim10^{-12}$ is required to avoid a too strong magnetic field for a stable solution with the MC effect.

\subsection{The dynamical equations of disk and corona with MC effects}

   The dynamical equations of the accretion flow are modified by considering the MC effects on the transfer of energy and angular momentum (Li \cite{li02a}; G09),
   \begin{equation}\label{5}
   \frac{\textup{d}}{\textup{d} r}(\dot{M}_{\textup{\scriptsize D}}L^{\dag}-g)=4\pi r(QL^{\dag}-H_{\textup{\scriptsize MC}}),
   \end{equation}
   \begin{equation}\label{6}
    \frac{\textup{d}}{\textup{d} r}(\dot{M}_{\textup{\scriptsize D}}E^{\dag}-g\Omega_i)=4\pi r(QE^{\dag}-H_{\textup{\scriptsize MC}}\Omega_i),
   \end{equation}
   where $Q$ and $g$ are respectively the viscously dissipated energy per unit disk surface and interior viscous torque of the disk, and $Q$ consists of two parts,
   \begin{equation}\label{7}
   Q=Q_{\textup{\scriptsize d}}+Q_{\textup{\scriptsize cor}},
   \end{equation}
   where $Q_{\textup{\scriptsize d}}$ is radiated from disk as black body,
   \begin{equation}\label{8}
    Q_{\textup{\scriptsize d}}=\sigma T^4_{\textup{\scriptsize eff}},
   \end{equation}
   and $Q_{\textup{\scriptsize cor}}$ is transported into the corona to heat it by magnetic reconnection of tangled small-scale magnetic fields, and it reads (Liu et al. \cite{liubf})
   \begin{equation}\label{9}
    Q_{\textup{\scriptsize cor}}=\frac{B^2_{\textup{\scriptsize D}}}{4\pi}V_{\textup{\scriptsize A}}=\frac{B^3_{\textup{\scriptsize D}}}{4\pi\sqrt{4\pi\rho}},
   \end{equation}
   where $B_{\textup{\scriptsize D}}$, $V_{\textup{\scriptsize A}}$ and $\rho$ are the intensity of the small-scale magnetic fields, the Alfven speed, and mass density, respectively.

   The quantity $H_{\textup{\scriptsize MC}}$ in Eqs. (5) and (6) is the flux of angular momentum and it reads
   \begin{equation}\label{10}
   H_{\textup{\scriptsize MC}}\equiv\frac{1}{4\pi r}\frac{\textup{d}T_{\textup{\scriptsize MC}}}{\textup{d} r}=\frac{1}{2\pi r\textup{d} r}\left(\frac{\textup{d}\Psi}{2\pi}\right)^2\frac{\Omega_i-\Omega_{i+1}}{\textup{d}Z_i},
   \end{equation}
   where $\textup{d}T_{\textup{\scriptsize MC}}$ and $\textup{d}\Psi$ are the torque exerted on to an infinitesimal annulus of the disk and magnetic flux threading the infinitesimal annulus, respectively.

   We can calculate a variety of magnetic transfer of energy in BH accretion disk by using an equivalent circuit in an analogous way to Macdonald \& Thorne (\cite{macdonald}) and W02 as shown in Fig.4. Three types of MC (MCDD, MCPD and MCHD) indicated in Fig.1 are given in Fig.4, in which a series of loops correspond to the adjacent magnetic surfaces arising from the rotation of the closed field lines.

   \begin{figure}
   \centering
    {\includegraphics[width=8.0cm]{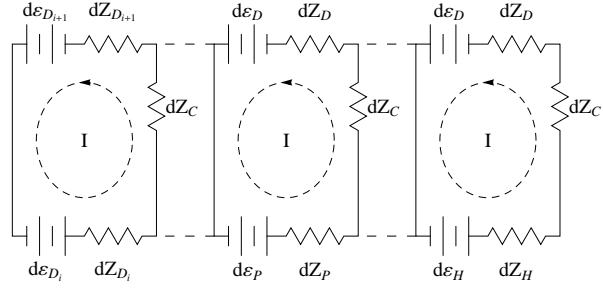}}
    \caption{An equivalent circuit for MCDD, MCPD and MCHD given from left to right.}\label{4}
   \end{figure}

   The quantities $\textup{d}Z_{\textup{\scriptsize D}}$, $\textup{d}Z_{\textup{\scriptsize P}}$, $\textup{d}Z_{\textup{\scriptsize H}}$ and $\textup{d}Z_{\textup{\scriptsize C}}$ are the resistances corresponding to the two adjacent magnetic surfaces in the disk, the plunging region, the BH horizon and the corona, respectively. The quantities $\textup{d}\varepsilon_{\textup{\scriptsize D}}$ and $\textup{d}\varepsilon_{\textup{\scriptsize H}}$ are respectively the electromotive forces generated by the rotation of the disk and the BH horizon as given by W02, and $\textup{d}\varepsilon_{\textup{\scriptsize P}}$ is the electromotive force due to the rotation of the plunging region.

   The quantities $\Omega_i$, $\Omega_{i+1}$ and $\textup{d}Z_i$ in Eqs. (6) and (10) are explained for the three different disk regions as follows.

   (1) \textbf{MCDD region:} $\Omega_i$ and $\Omega_{i+1}$ are respectively the angular velocities of the inner footpoint $i$ and outer footpoint $i+1$ of the closed field line on the disk, and the resistance $\textup{d}Z_i$ is defined by
   \begin{equation}\label{11}
    \textup{d}Z_i=\textup{d}Z_{\textup{\scriptsize cor}}=\frac{R_{\textup{\scriptsize cor}}\textup{d}r}{2\pi r},
   \end{equation}
   where $R_{\textup{\scriptsize cor}}$ is the average areal resistivity of the disk and corona, which is related to $R_{\textup{\scriptsize H}}$, the surface resistivity of the BH horizon as follows,
   \begin{equation}\label{11}
   \left\{
   {\begin{array}{l}
   R_{\textup{\scriptsize cor}}=\eta_{\textup{\scriptsize R}} R_{\textup{\scriptsize H}} \\
   R_{\textup{\scriptsize H}}=4\pi=377\textup{ohm},
   \end{array}} \right.
   \end{equation}
   where $\eta_{\textup{\scriptsize R}}$ is a parameter to adjust the value of $R_{\textup{\scriptsize cor}}$. For simplicity, we set $\eta_{\textup{\scriptsize R}}=0.1$ in calculations.

   (2) \textbf{MCHD region:} $\Omega_i=\Omega_{\textup{\scriptsize H}}$ and $\Omega_{i+1}=\Omega_{\textup{\scriptsize D}}$ are the angular velocities of the BH horizon and the disk respectively. The resistance
   \begin{equation}\label{13}
   \textup{d}Z_i=\textup{d}Z_{\textup{\scriptsize H}}+\textup{d}Z_{\textup{\scriptsize cor}},
   \end{equation}
   where
   \begin{equation}\label{14}
   \textup{d}Z_{\textup{\scriptsize H}}=2\rho_{\textup{\scriptsize H}}\textup{d}\theta/\varpi.
   \end{equation}
   The parameters in Eq. (14) are given by
   \begin{eqnarray}\label{15}
   \varpi=(\Sigma_{\textup{\scriptsize H}}/\rho_{\textup{\scriptsize H}})\sin\theta, \ \ \ \ \ \ \rho_{\textup{\scriptsize H}}^2\equiv r^2_{\textup{\scriptsize H}}+a^2\cos^2\theta,  \nonumber \\
   \Sigma_{\textup{\scriptsize H}}\equiv 2Mr_{\textup{\scriptsize H}}, \ r_{\textup{\scriptsize H}}=M(1+q), \ q=\sqrt{1-a^2_*}.
   \end{eqnarray}

   (3) \textbf{MCPD region:} $\Omega_i=\Omega_{\textup{\scriptsize P}}$ and $\Omega_{i+1}=\Omega_{\textup{\scriptsize D}}$ are respectively the angular velocities of the plunging region and the disk, and $\Omega_{\textup{\scriptsize P}}$ is given as follows (Shapiro \& Teukolsky \cite{Shapiro}; Wang \cite{wang}),
   \begin{equation}\label{16}
   \Omega_{\textup{\scriptsize P}}=\frac{(r-2M)L^\dag+2aME^\dag}{(r^3+a^2r+2Ma^2)E^\dag-2aML^\dag}.
   \end{equation}
   The resistance is the same as that in MCDD expressed by Eq. (11).

   We adopt the same assumption as given by G09 about the corona: the optical depth of the corona $\tau_{\textup{\scriptsize cor}}=1$ and the height of the corona $l=10r_{\textup{\scriptsize ms}}$.

\section{ Fitting the QPO frequencies and X-ray spectra  }

   In our model, the toroidal electric current interacts with the dynamics of the accretion disk. To solve the dynamic Eqs. (5) and (6), we must know the configuration of the large-scale magnetic fields generated by the toroidal electric current, whose distribution is related to the surface density of the disk matter by Eq. \eqref{3}, and $\rho_{\textup{\scriptsize m}}$ is in turn figured out by solving the dynamical equations. For simplicity, we assume that initially there is no electric current in the disk, and solve the Eqs. (5) and (6) with $H_{\textup{\scriptsize MC}}=0$ to get the global solution of the disk-corona system. The emerged spectrum of the disk-corona system is then simulated with Monte Carlo method based on the code developed by G09. The free parameters of the disk-corona model, e.g., BH spin $a_*$ and mass accretion rate $\dot{m}$, can be determined by fitting the observed spectrum. So the surface density of the disk matter is obtained. Then we consider the interaction between the electric current and disk-corona with the iterative algorithm which consists of the following steps:

(i) Assuming a value of the power law index $n$ in Eq. \eqref{eq3}, e.g. $n=5$.

(ii) Calculating the surface density of the toroidal electric current in the disk and the configuration of the large-scale magnetic fields.

(iii) Resolving the disk-corona system by taking the MC effects into account.

(iv) Repeating steps (ii) and (iii) until the surface density of disk matter remains stable.

(v) Calculating the frequency $\nu_{\textup{\scriptsize QPO}}$ by Eq. \eqref{4}.

(vi) Repeating steps (i)-(v) until the QPO frequency is in accordance with the observations.

Steps (ii)-(iii) should be repeated several times before the surface density of disk matter becomes stable. Finally, we simulate the emerged spectrum again until it is unchanged.

To avoid the negative radiation flux from the inner disk due to the transfer of energy and angular momentum to the outer disk by the MCDD as argued by Gan et al. (\cite{gan07}), we adjust the radius of the inner boundary of the disk to deviate outwards from ISCO. Although the deviation is less than 10\% of the radius of ISCO, it results in $\sim$30\% decrease of QPO frequency.

\subsection{Fitting the HFQPOs in XTE J1859+226 and XTE J1650$-$500 with SPL state}

Observations show that HFQPOs in BHBs are associated with the SPL state which is characterized by high luminosity (therefore high accretion rate), strong power-law component and the steep power-law index ($\Gamma>2.4$). We fit the single HFQPOs and the corresponding X-ray spectra of two BHBs XTE J1859+226 and XTE J1650-500. The comparisons between the simulated spectra and the observed ones are shown for XTE J1859+226 and XTE J1650$-$500 in Figs. 5 and 6, respectively. The total spectra consist of the thermal component emitted from the disk and the power-law component generated by the inverse Compton scattering of the soft photons by the relativistic electrons in the corona.

The fitting parameters are listed in Table 1, in which $\dot{m}$ and $\alpha$ are respectively the mass accretion rate in terms of the Eddington accretion rate ($1.4\times10^{18} m_{\textup{\scriptsize BH}}$ $\textup{g s}^{-1}$) and the viscosity parameter. The BH mass of XTE J1859+226 was estimated in the range 7.6$-$12$M_\odot$ (MR06). We fit the spectrum of XTE J1859+226 observed in October 16$-$18 of 1999 (MR06) with the BH mass of the upper limit 12$M_\odot$ and of the lower limit 7.6$M_\odot$. Since the BH mass of XTE J1650$-$500 has not been well estimated, we take a larger mass 7.3$M_\odot$ and a smaller one 4$M_\odot$, which are estimated by Orosz et al. (\cite{orosz}) to fit the spectrum. Very large BH spin is needed to fit the spectra for the lower limit to the BH mass, and the lower limit to BH spin corresponding to each BH mass is given in Table 1. The spectrum of XTE J1650$-$500 is fitted with the lower and upper limits to spin for each BH mass, while the spectrum of XTE J1859+226 can be fitted only for an extreme Kerr BH with the lower limit to the BH mass 7.6$M_\odot$. The value of the viscosity parameter $\alpha$ is selected properly in the range 0.1$-$0.3 to fit the spectrum of each source. The difference of the value arises probably from the difference of the strength of the small-scale magnetic field in the disk if the viscous process is dominated by the tangled small-scale magnetic field. The values of the fitting parameters $n$ and $\eta$ for QPO frequencies are listed in the last two column in Table 1. We fix $\eta$ at a certain value for each scale of BH mass. BH systems with smaller mass have stronger magnetic field allowing of larger value of $\eta$. A larger value of $n$ is needed with a smaller BH mass, implying that the toroidal electric currents are concentrated very close to ISCO.

   \begin{figure}
   \centering
    \includegraphics[width=4.9cm]{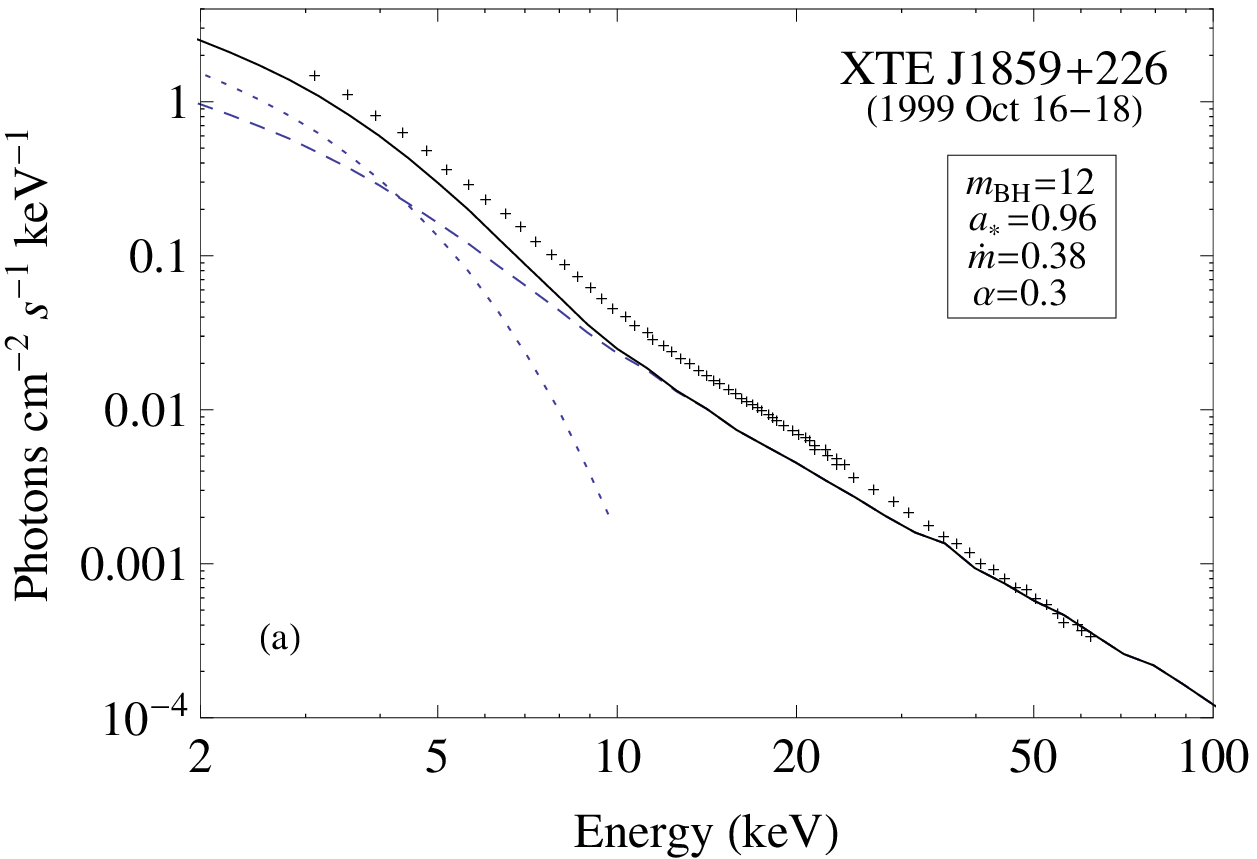}
    \includegraphics[width=4.9cm]{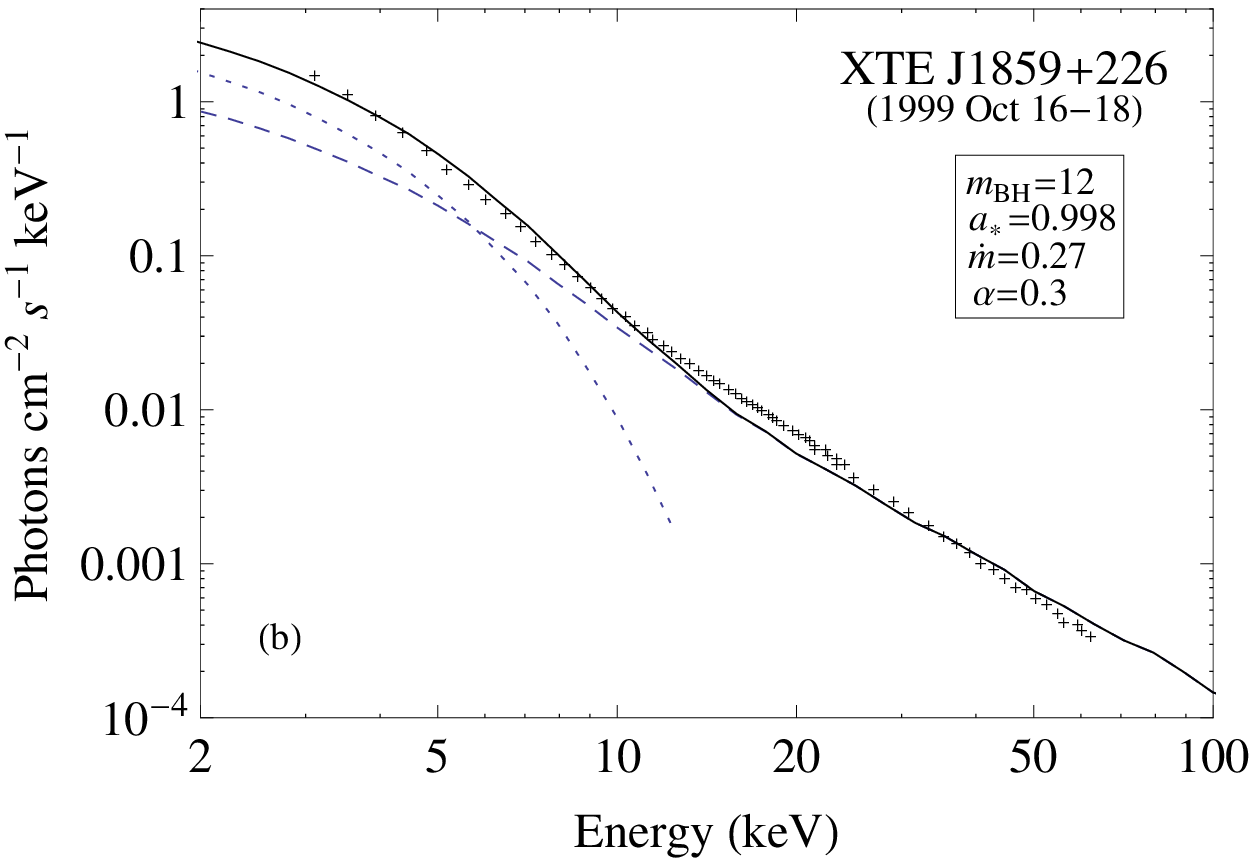}
    \includegraphics[width=4.9cm]{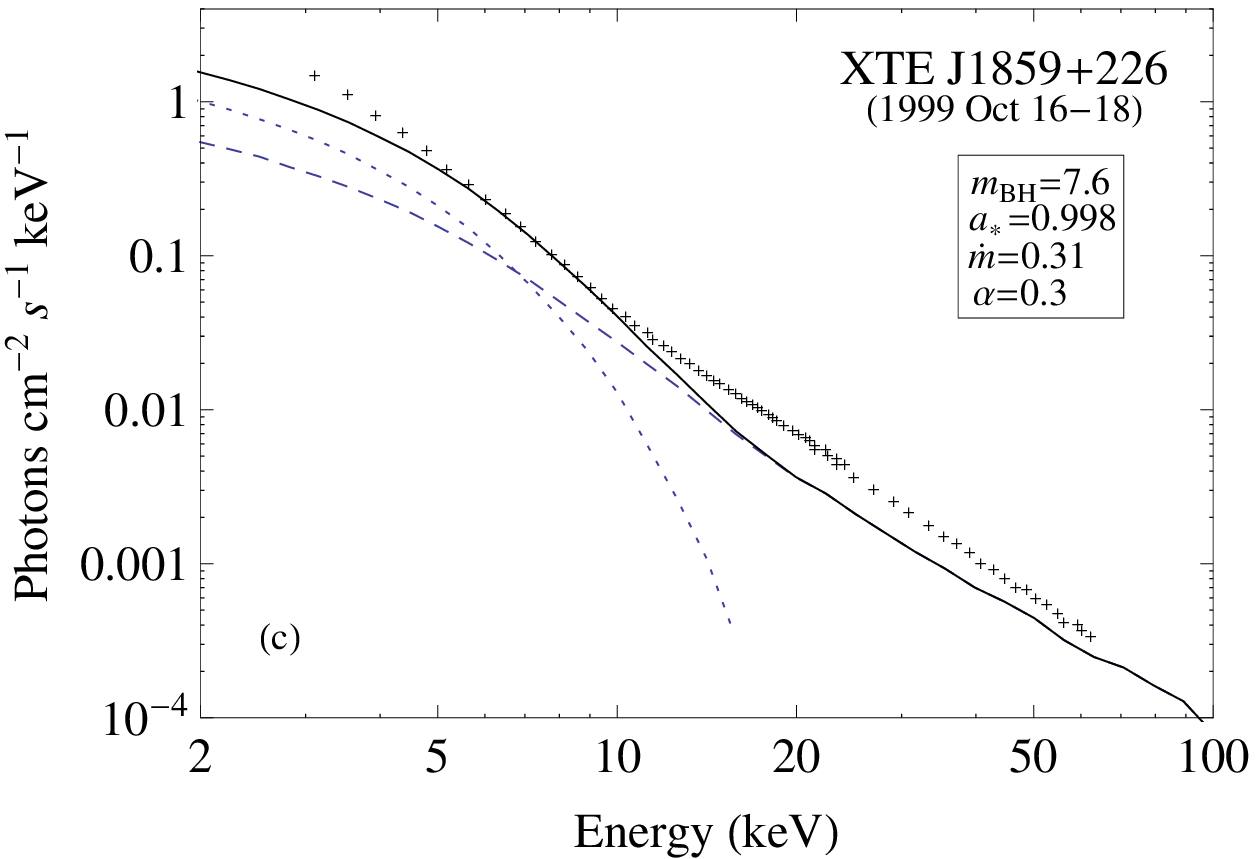}
    \caption{The simulated spectra of XTE J1859+226 with different parameters:
 (a) $m_{\textup{\scriptsize BH}}=12$, $a_*=0.96$, $\dot{m}=0.38$ and
 $\alpha=0.3$; (b) $m_{\textup{\scriptsize BH}}=12$, $a_*=0.998$, $\dot{m}=0.27$ and $\alpha=0.3$;
 (c) $m_{\textup{\scriptsize BH}}=7.6$, $a_*=0.998$, $\dot{m}=0.31$, and
  $\alpha=0.3$. The total spectrum and its thermal and Comptonized components are plotted in solid, dotted and dashed lines, respectively. The observation data are taken from MR06. The source distance is set at 11 kpc (Zurita et al. \cite{zurita}; MR06) and the inclination $i=70^\circ$ is assumed.}\label{5}
   \end{figure}

   \begin{figure}
   \centering
    \includegraphics[width=6cm]{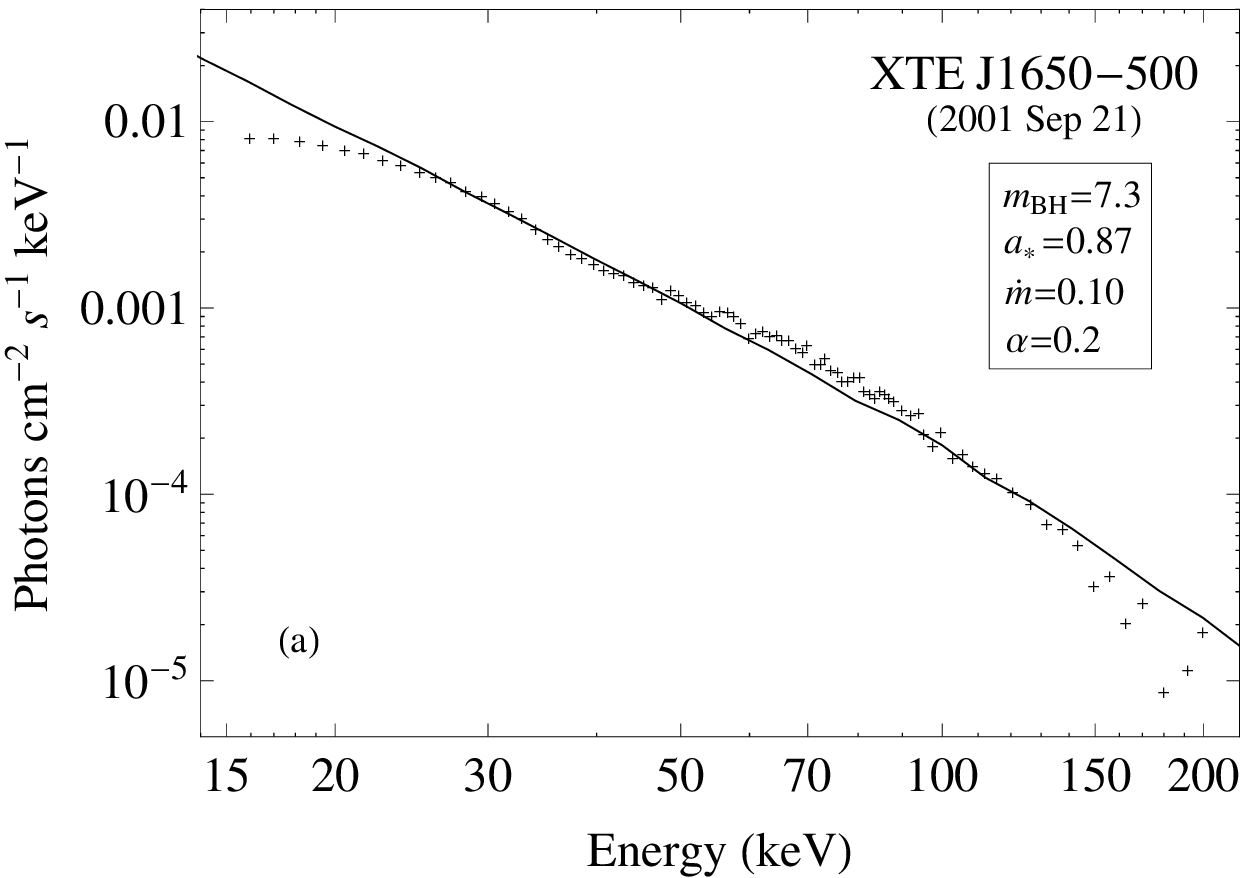}  \ \ \ \ \ \  \includegraphics[width=6cm]{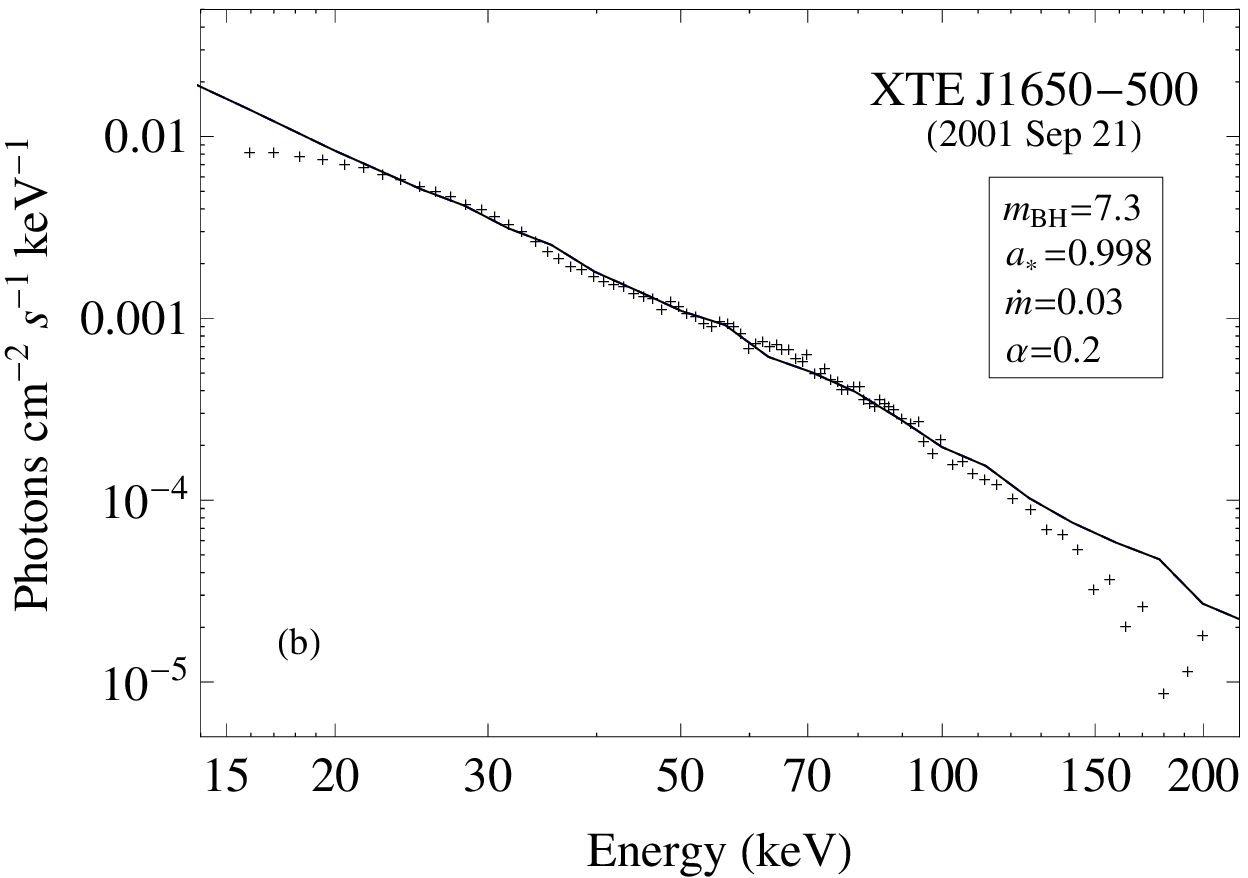} \\
    \includegraphics[width=6cm]{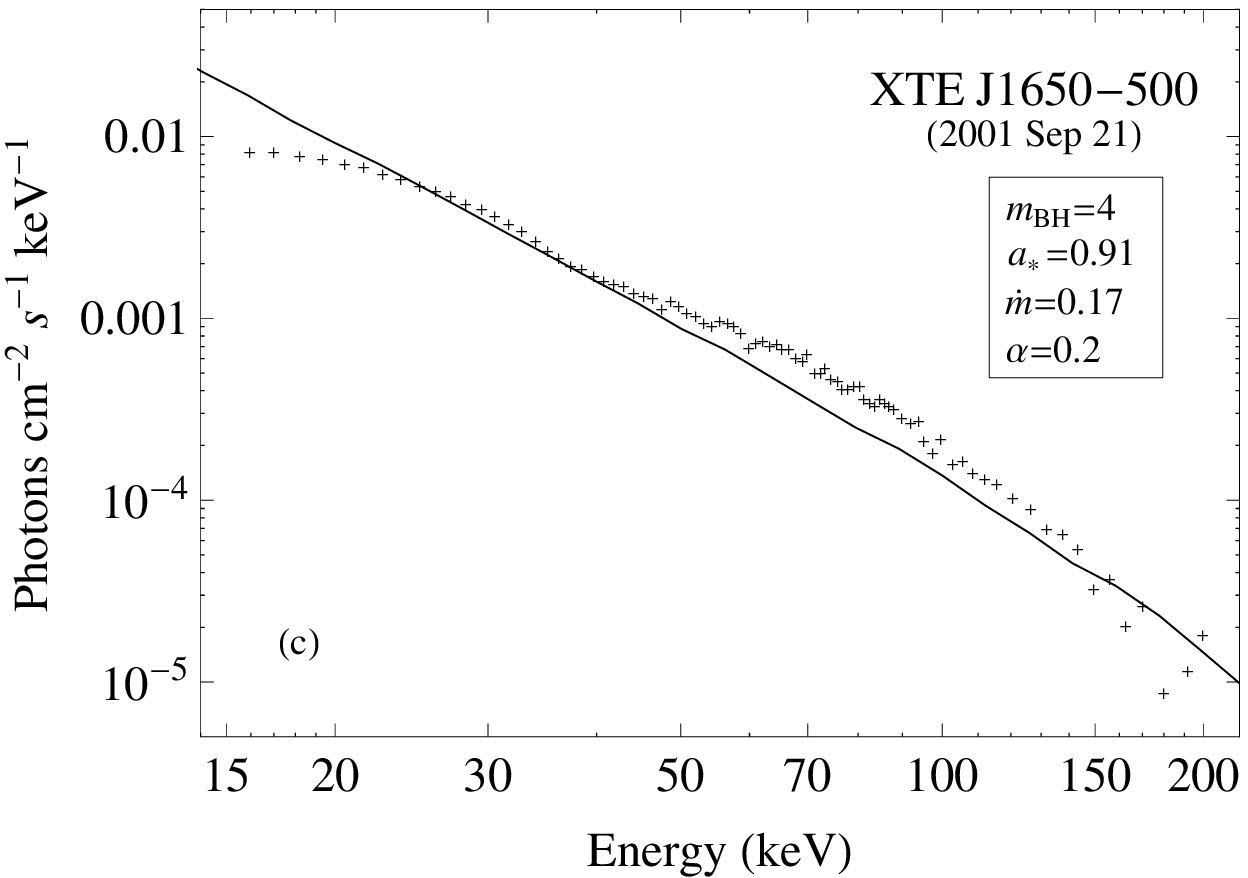}  \ \ \ \ \ \  \includegraphics[width=6cm]{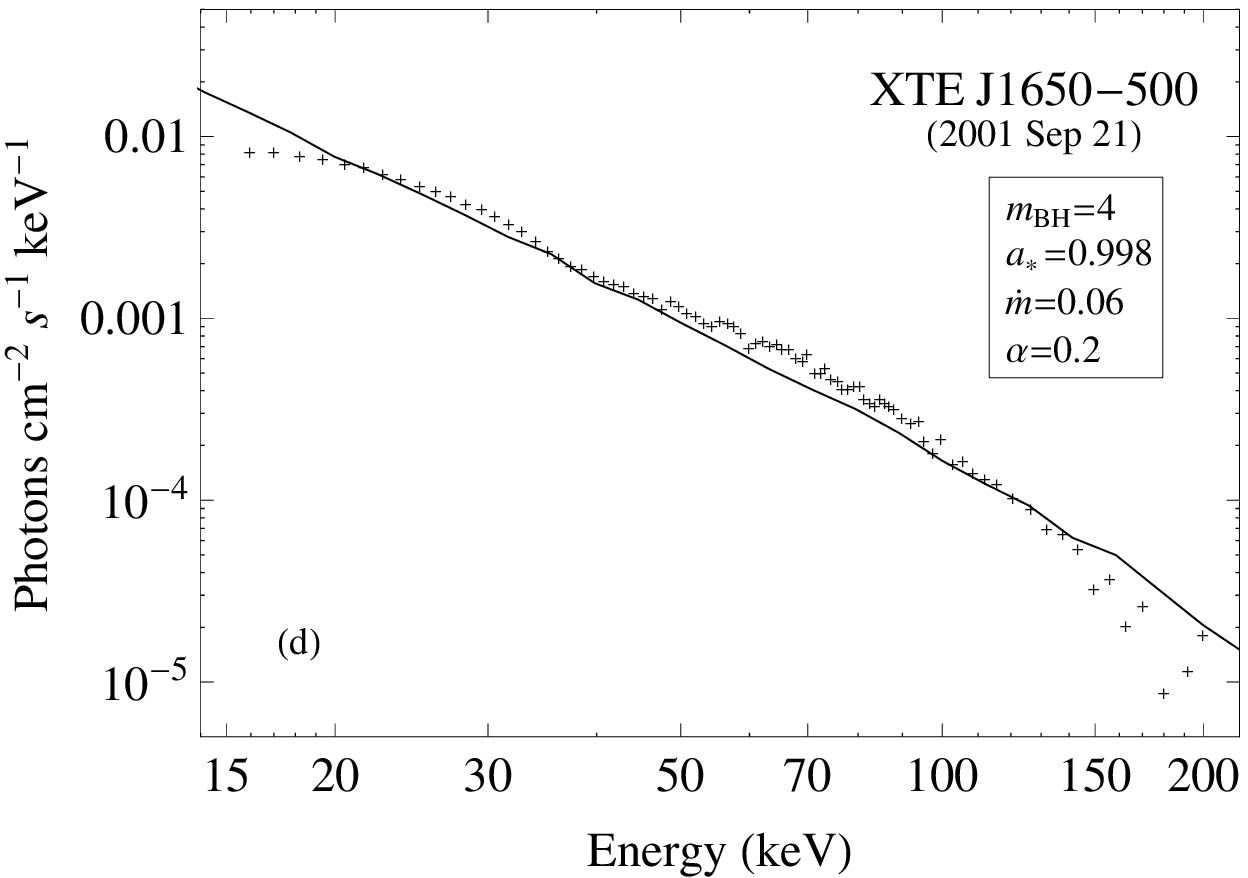}
    \caption{The simulated spectra of XTE J1650$-$500 with different parameters:
 (a) $m_{\textup{\scriptsize BH}}=7.3$, $a_*=0.87$, $\dot{m}=0.10$ and
 $\alpha=0.2$; (b) $m_{\textup{\scriptsize BH}}=7.3$, $a_*=0.998$, $\dot{m}=0.03$ and $\alpha=0.2$;
 (c) $m_{\textup{\scriptsize BH}}=4$, $a_*=0.91$, $\dot{m}=0.17$, and
  $\alpha=0.2$; (d) $m_{\textup{\scriptsize BH}}=4$, $a_*=0.998$, $\dot{m}=0.06$, and
  $\alpha=0.2$. The observation data are taken from Miniutti et al. (\cite{miniutti}). The source distance is set at 2.6 kpc (Homan et al. \cite{homan06}) and the inclination $i=70^\circ$ is assumed.}\label{6}
   \end{figure}

   \begin{table}
 \centering
 \begin{minipage}{110mm}
  \caption{The fitting parameters for QPO frequencies and X-ray spectra.}
  \begin{tabular}{@{}lccccccc@{}}
  \hline
   Source & $\nu_{\textup{\scriptsize QPO}}$(Hz) & $m_{\textup{\scriptsize BH}}$ & $a_*$ & $\dot{m}$ & $\alpha$ & $n$ & $\eta$ \\
 \hline
 XTE J1859+226 & 190$^a$ & 12$^b$ & 0.96 & 0.38 & 0.3 & 8.4\\
  & & & 0.998 & 0.27 & 0.3 & 8.9 \\
  \cline{3-7}
  & & 7.6$^b$ & 0.998 & 0.31 & 0.3 & 14.6 \\
  \cline{1-7}
  XTE J1650$-$500 & 250$^c$ & 7.3$^d$ & 0.87 & 0.10 & 0.2 & 3.1\\
  & & & 0.998 & 0.03 & 0.2 & 9.1 & $10^{-12}$\\
  \cline{3-7}
  & & 4$^d$ & 0.91  & 0.17 & 0.2 & 17.7 \\
  & &   & 0.998 & 0.06 & 0.2 & 23.1 \\
  \cline{1-7}
 GRS 1915+105 & 67$^e$ & 18$^b$ & 0.998 & 0.25 & 0.14 & 16.9 \\
              &    & 10$^b$ & 0.998 & 0.40 & 0.16 & 27.0 \\
 \hline
 NGC 5408 X-1 & 0.02$^f$ & $1.0\times10^5$ & 0.95 & 0.012 & 0.1 & 15.1 & $10^{-16}$\\
              &          &      & 0.998 & 0.005 & 0.1 & 15.5 \\
\hline
RE J1034+396 & 0.00027$^g$ & $7.0\times 10^{6h}$ & 0.99 & 0.15 & 0.3 & 16.4 & $10^{-17}$\\
             &             &                     & 0.998 & 0.14 & 0.3 & 16.4 \\
\cline{3-8}
             &             & $2.0\times 10^{6h}$ & 0.998 & 0.13 & 0.3 & 39.7 & $10^{-15}$\\
\hline
\end{tabular}
\newline
\newline $^{a}$Cui et al. (\cite{cui}); $^{b}$MR06; $^{c}$Homan et al. (\cite{homan03}); $^{d}$Orosz et al. (\cite{orosz}); $^{e}$Morgan et al. (\cite{morgan}); $^{f}$Strohmayer et al. (\cite{strohmayer07}); $^{g}$Gierlinski et al. (\cite{gierlinski}); $^{h}$Zhou et al. (\cite{zhou})
\end{minipage}
\end{table}

\subsection{QPOs in NGC 5408 X-1 and RE J1034+396}

One of attractive features of our model is that it is applicable to fit QPOs with corresponding X-ray spectra observed in the BH systems of different scales. Strohmayer et al. (\cite{strohmayer07}) discovered a strong 20 mHz QPO in the ULX NGC 5408 X-1, and the X-ray timing and spectral properties are analogous to those exhibited by Galactic stellar-mass BHs in the `very high' or SPL state. The fitting parameters for the QPO and X-ray spectrum of NGC 5408 X-1 are listed also in Table 1. The simulated spectra for different parameters and the comparisons to the observed energy spectrum are shown in Fig. 7. However, at present, the BH masses in the ULXs are constrained to a wide range from $10^2M_\odot$ to $10^5M_\odot$ (Miller et al. \cite{miller}; Cropper et al. \cite{cropper}; Roberts et al. \cite{roberts}; Wu \& Gu \cite{wu}). We take a very large BH mass $10^5M_\odot$ to fit the data of NGC 5408 X-1 since both the X-ray spectrum and QPO frequency are fitted better for larger BH mass. The QPO frequency and spectrum are both taken from Strohmayer et al. (\cite{strohmayer07}) with the same Hydrogen column density, which are fitted in our model with the lower and upper limits to the BH spin as shown in Table 1.

   \begin{figure}
   \centering
    \includegraphics[width=6cm]{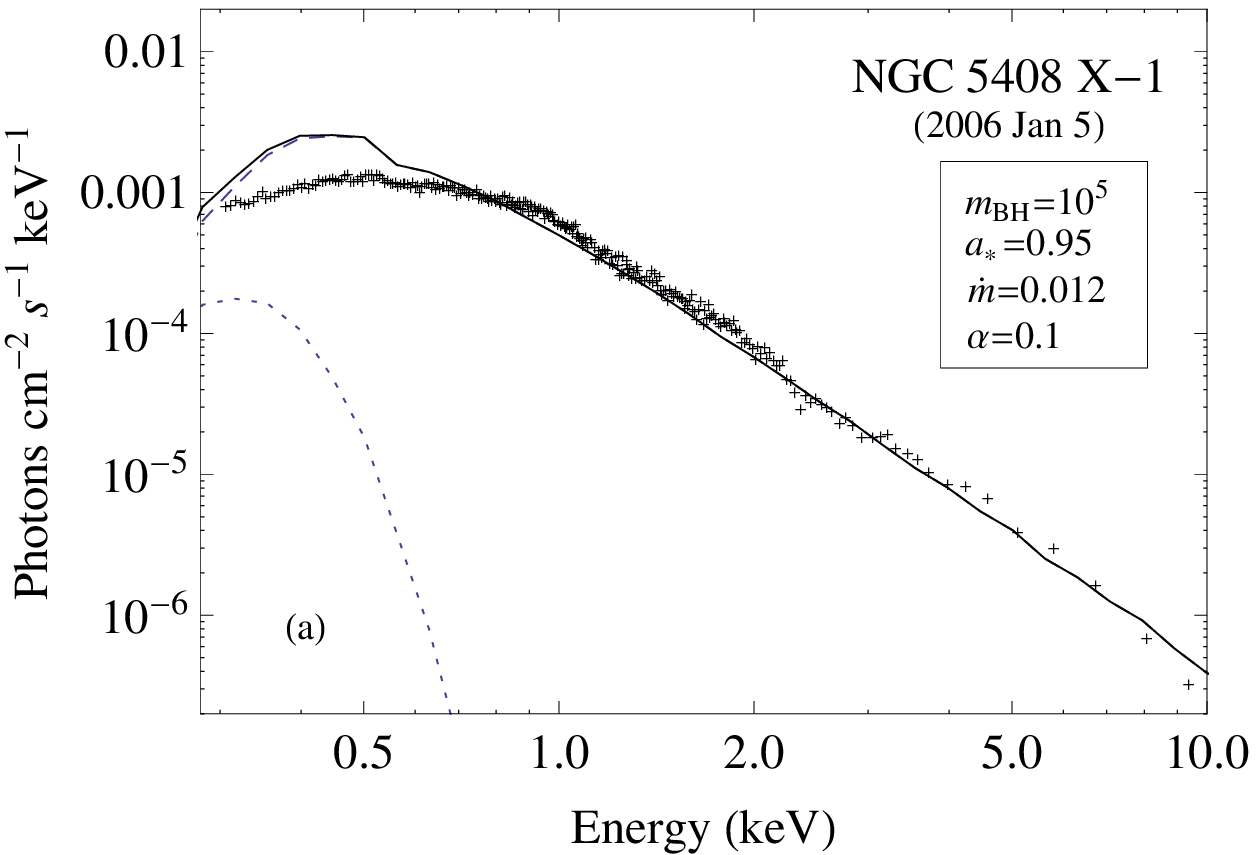} \ \ \ \ \ \
    \includegraphics[width=6cm]{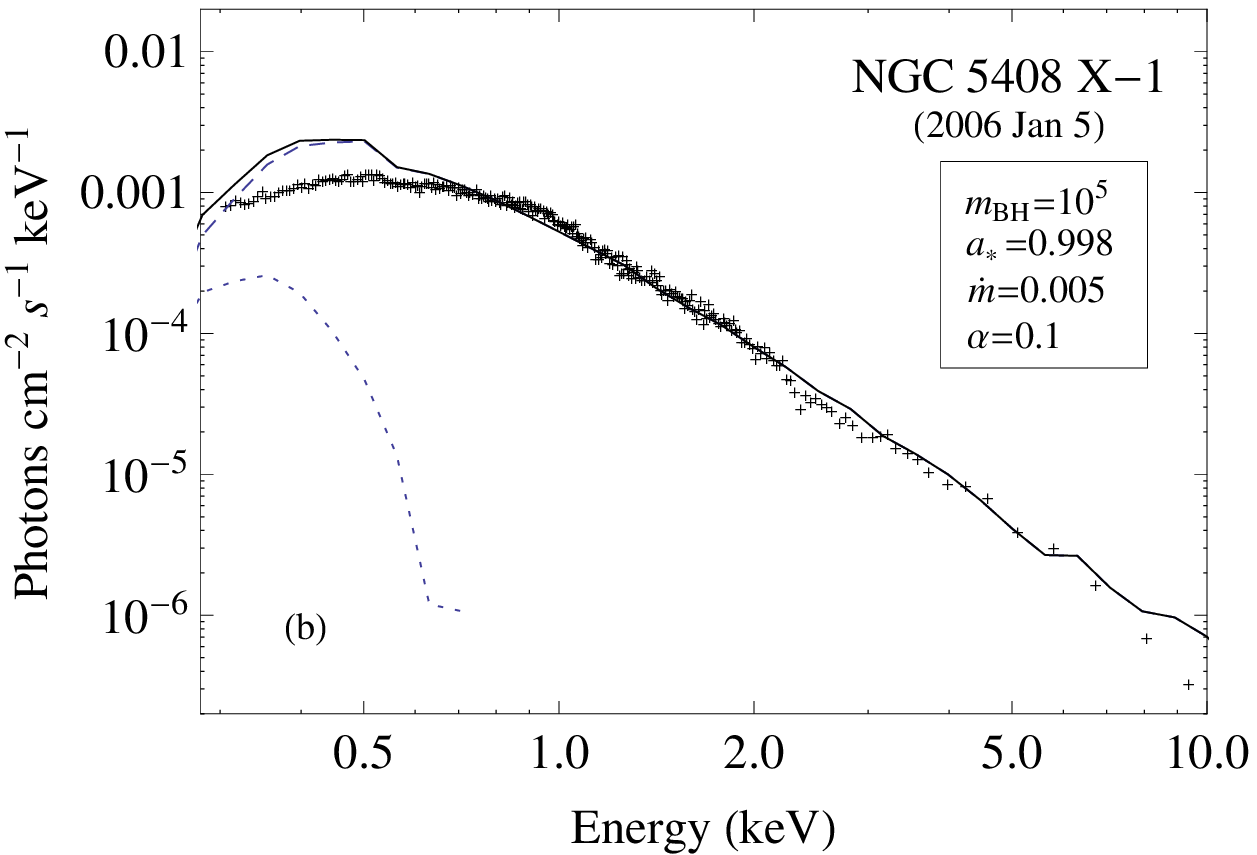}
    \caption{The simulated spectra of NGC 5408 X-1 with different parameters:
 (a) $m_{\textup{\scriptsize BH}}=10^5$, $a_*=0.95$, $\dot{m}=0.012$ and
 $\alpha=0.1$; (b) $m_{\textup{\scriptsize BH}}=10^5$, $a_*=0.998$, $\dot{m}=0.005$ and $\alpha=0.1$.
The plot style is the same as Figure 5. The observation data are taken from Strohmayer et al. (\cite{strohmayer07}). The source distance is set at 4.8 Mpc (Karachentsev et al. \cite{karachentsev}) and the inclination $i=75^\circ$ is assumed. The total Hydrogen column density is set as $n_{\rm H}=13\times10^{20}$ cm$^{-2}$ (Strohmayer et al. \cite{strohmayer07}).}\label{7}
   \end{figure}

   \begin{figure}
   \centering
    \includegraphics[width=6cm]{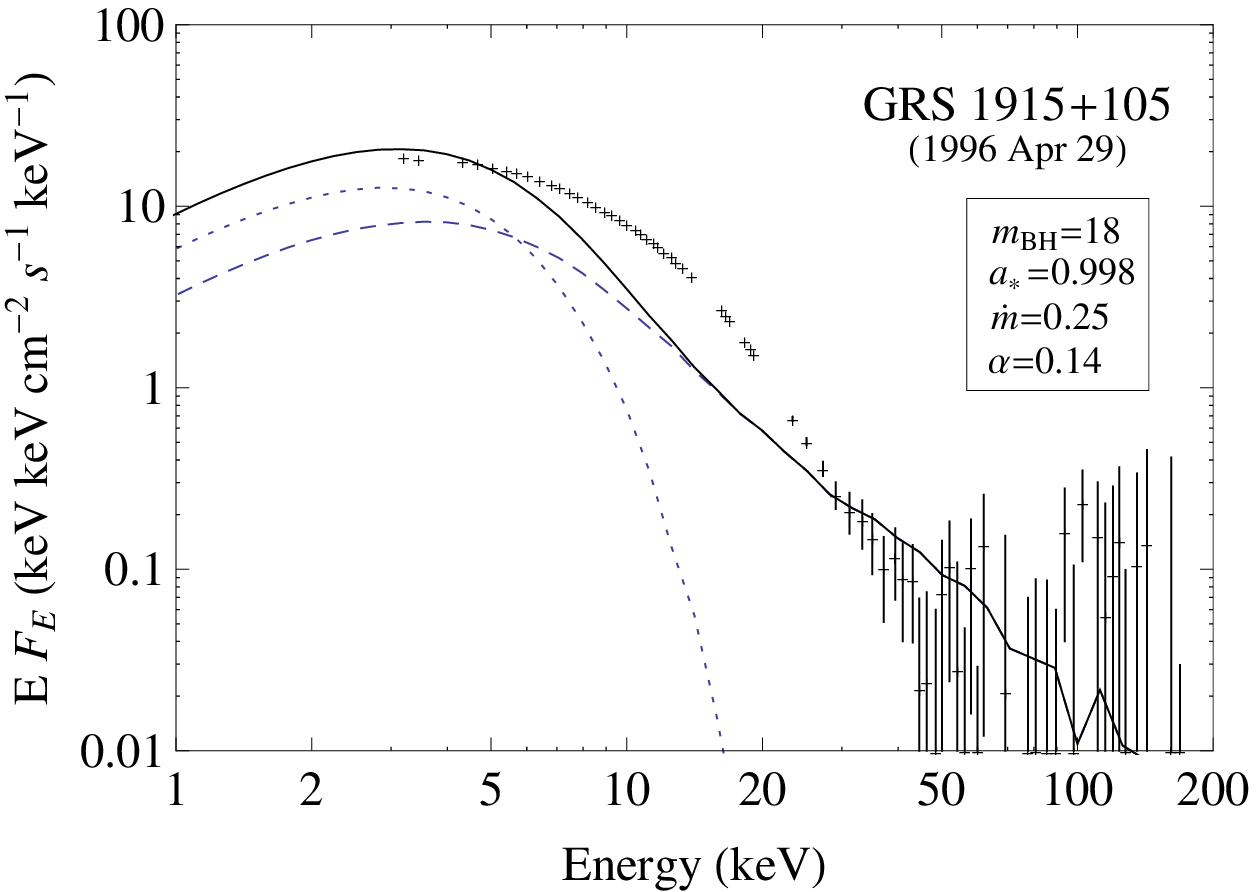} \ \ \ \ \ \
    \includegraphics[width=6cm]{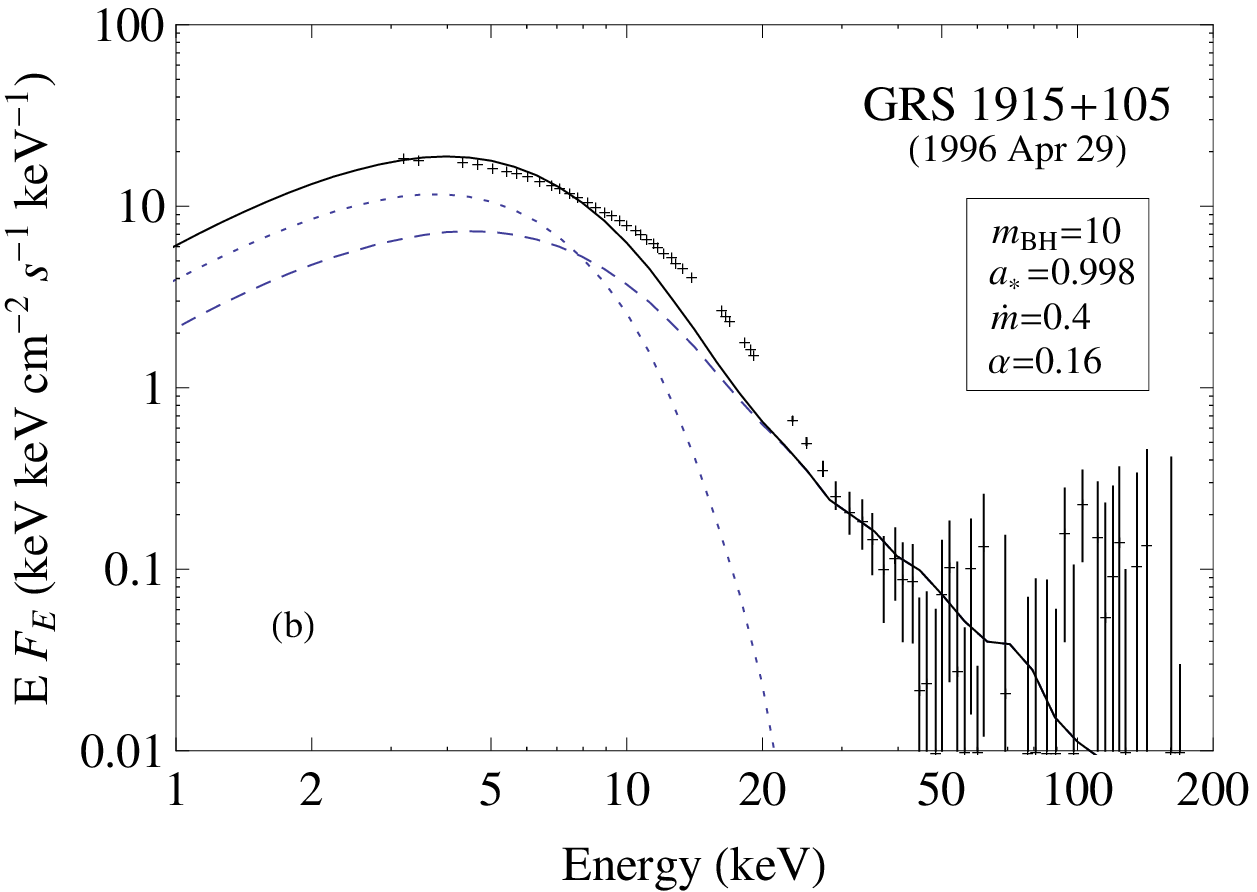}
    \caption{The simulated spectra of GRS 1915+105 with different parameters:
 (a) $m_{\textup{\scriptsize BH}}=18$, $a_*=0.998$, $\dot{m}=0.25$ and
 $\alpha=0.14$; (b) $m_{\textup{\scriptsize BH}}=10$, $a_*=0.998$, $\dot{m}=0.4$ and $\alpha=0.16$.
  The plot style is the same as Figure 5. The observation data are taken from Middleton \& Done (\cite{middleton10}). The source distance is set at 11 kpc (McClintock et al. \cite{mcclintock}) and the inclination $i=66^\circ$ (Fender et al. \cite{fender}) is used.}\label{8}
   \end{figure}

   \begin{figure}
   \centering
    \includegraphics[width=4.9cm]{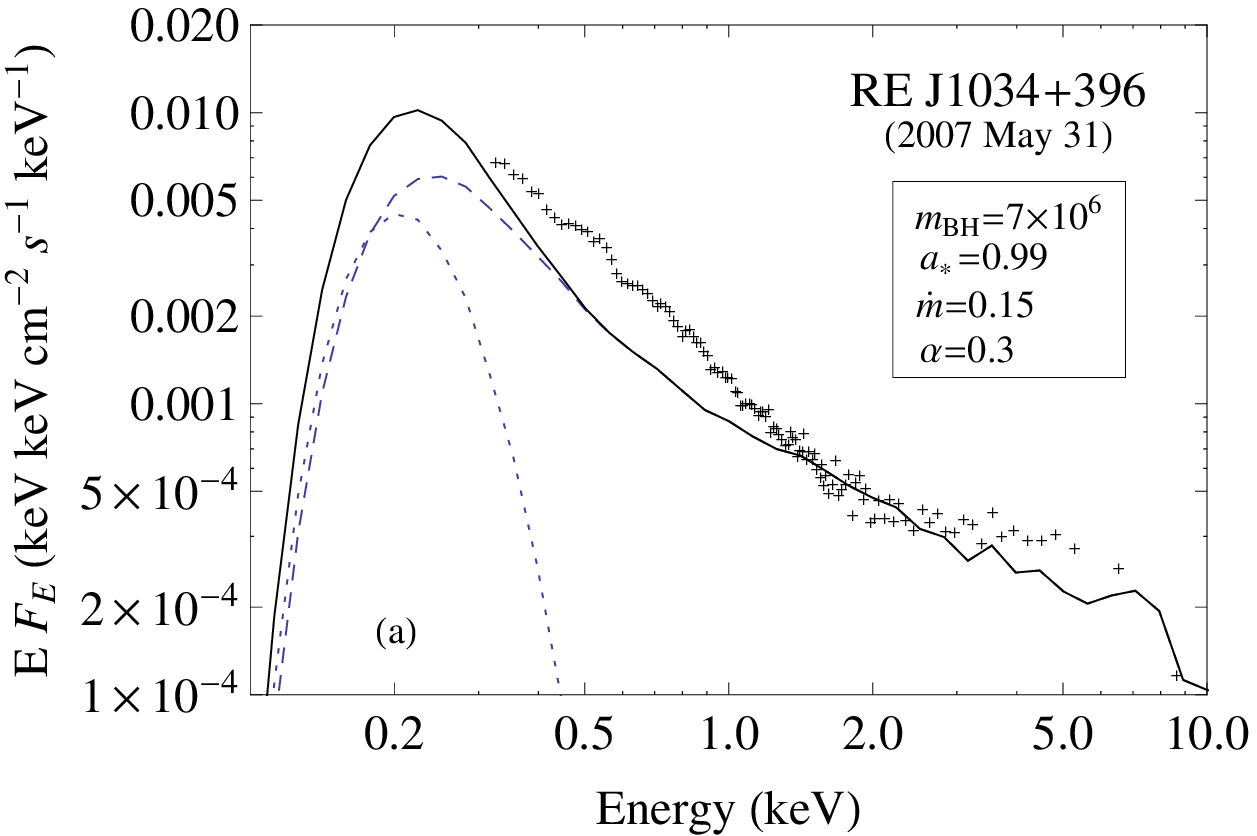}
    \includegraphics[width=4.9cm]{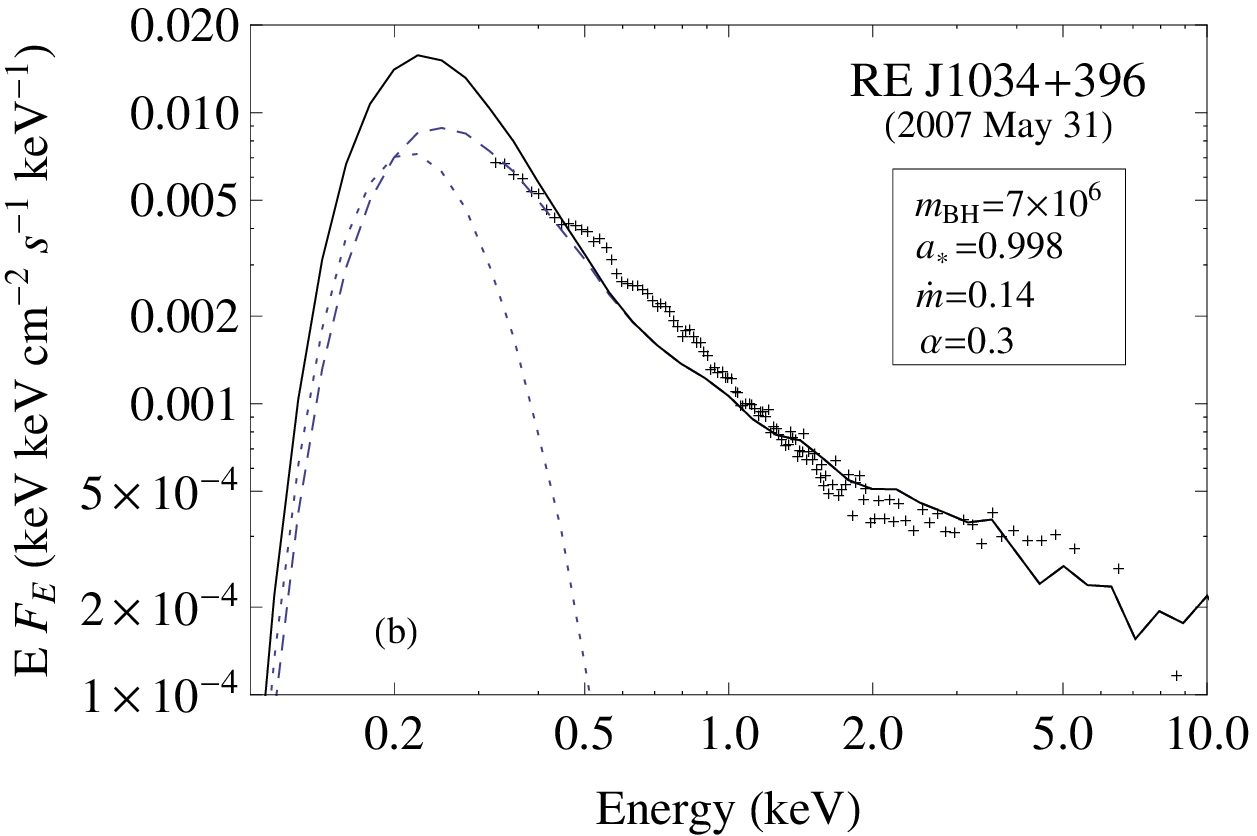}
    \includegraphics[width=4.9cm]{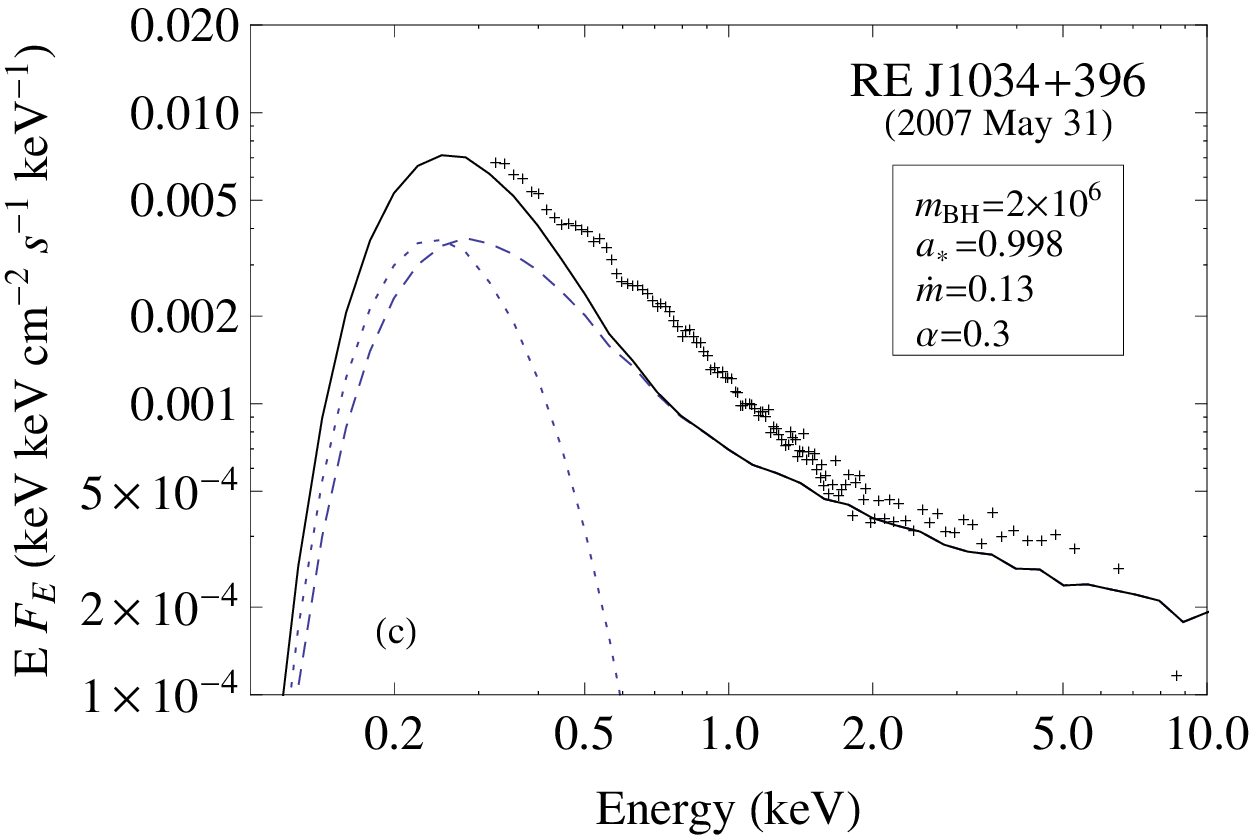}
    \caption{The simulated spectra of RE J1034+396 with different parameters:
 (a) $m_{\textup{\scriptsize BH}}=7\times10^6$, $a_*=0.99$, $\dot{m}=0.15$ and
 $\alpha=0.3$; (b) $m_{\textup{\scriptsize BH}}=7\times10^6$, $a_*=0.998$, $\dot{m}=0.14$ and $\alpha=0.3$;
 (c) $m_{\textup{\scriptsize BH}}=2\times10^6$, $a_*=0.998$, $\dot{m}=0.13$, and
  $\alpha=0.3$. The plot style is the same as Figure 5. The observation data are taken from Middleton et al. (\cite{middleton09}). The source distance is set at 125.9 Mpc ($z$=0.042, $H_0$=100 km s$^{-1}$ Mpc$^{-1}$) and the inclination $i=40^\circ$ is assumed. We only consider the minimum galactic absorption fixed at $1.31\times10^{20}$ cm$^{-2}$ (Middleton et al. \cite{middleton09}).}\label{9}
   \end{figure}

The first convincing QPO of AGNs was reported by Gierlinski et al. (\cite{gierlinski}) in RE J1034+396. Middelton \& Done (\cite{middleton10}) suggests that the QPO discovered in RE J1034+396 has an analogy to the 67 Hz QPO seen in the BHB GRS 1915+105 due to their similar `hot disk dominated' energy spectra. Unlike other HFQPOs, the 67 Hz QPO in GRS 1915+105 is an exceptional case, which appears in thermal-dominant (TD) state (MR06). In this subsection, we fit both the 67 Hz QPO in GRS 1915+105 and the  0.00027Hz QPO in RE J1034+396 as a comparison. The spectrum of GRS 1915+105 showing the 67 Hz QPO is taken from Middleton \& Done (\cite{middleton10}). And the spectrum of RE J1034+396 showing the 0.00027Hz QPO is taken from Middleton et al. (\cite{middleton09}) with the same minimum galactic absorption fixed at $1.31\times10^{20}$cm$^{-2}$. The `hot disk dominated' spectra of GRS 1915+105 and RE J1034+396 are both fitted with almost maximum BH spins as shown in Table 1, and their comparisons with the observed spectra are shown in Figs 8 and 9, respectively. The disparities between the simulated spectra and the observed ones in the energy bands 10$-$30 keV for GRS 1915+105 and 0.5$-$1 keV for RE J1034+396 indicate probably that a second Comptonization process is needed to generate the TD spectrum as shown in Figs 10 and 11 of Middelton \& Done (\cite{middleton10}), where a low temperature, optically thick thermal Comptonization is added to fit the spectra. This Comptonization may be generated from the transition layer between the disk and the corona.

Inspecting Table 1 and Figs 5$-$9, we find that the QPOs and the corresponding X-ray spectra of BH systems of different scales can be fitted with the magnetic reconnection of the large-scale magnetic fields based on the disk-corona model. QPOs in massive BHs have similar features with those of BHBs, e.g. very centralized distribution of the electric currents or large value of $n$, and its inverse proportion to BH mass. They are either related to the SPL state, like XTE J1859+226 and XTE J1650$-$500, or related to the TD state, like GRS 1915+105. These similar characteristics hint probably that QPOs in BH systems of different scales may have the same origin and are associated with the same spectral state.

\section{Discussion}

   \begin{figure}[b]
   \centering
    \includegraphics[width=8cm]{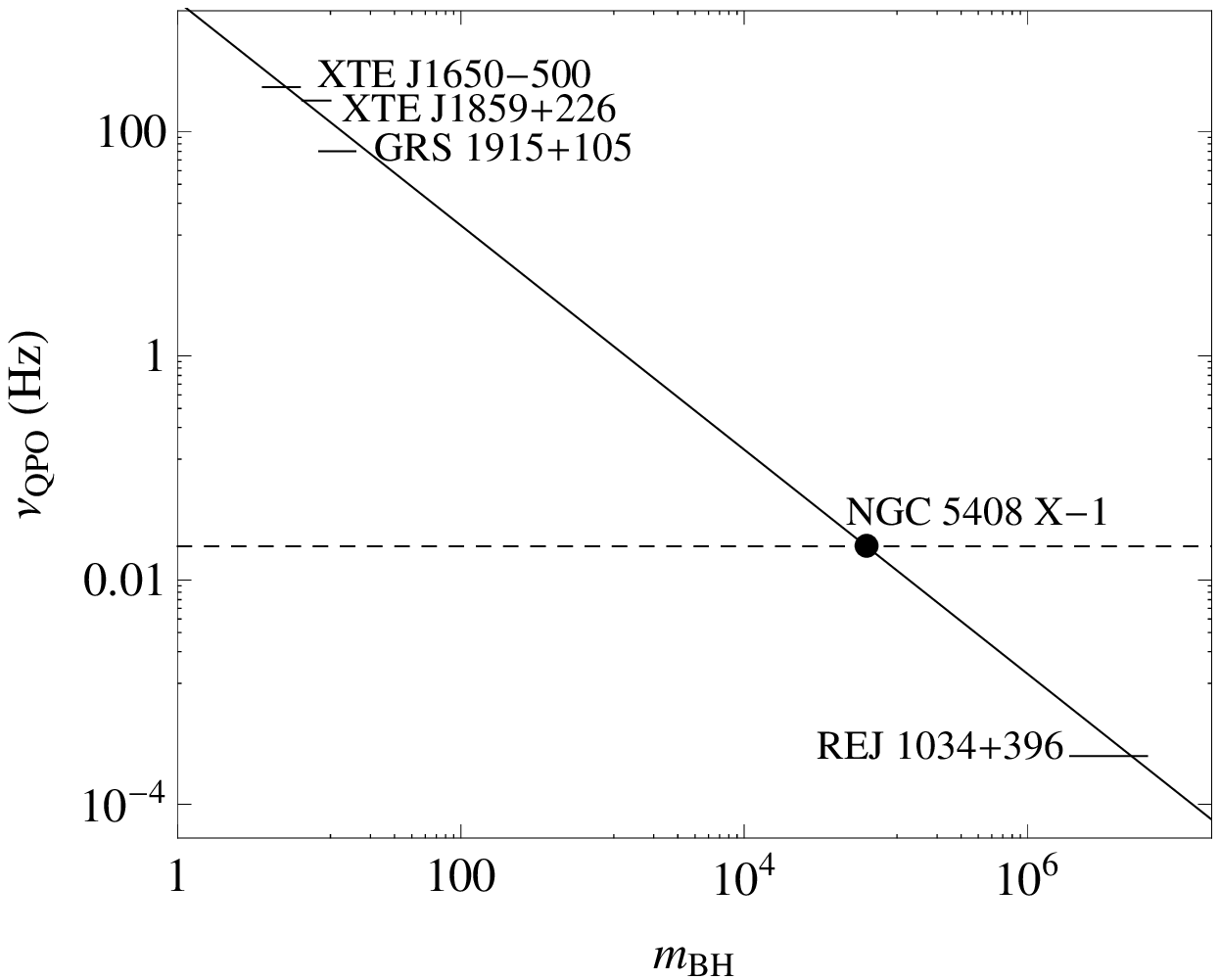}
    \caption{Relationship between the BH mass and the QPO frequency. The relation is fitted with $m_{\textup{\scriptsize BH}}=1460\nu^{-1}_{\textup{\scriptsize QPO}}$ shown by the skew line. The horizontal short lines represent the BH mass ranges of the sources. NGC 5408 X-1 is shown on the skew line (black dot) corresponding to the 0.02 Hz QPO (dashed line).}\label{10}
   \end{figure}

In this paper, a toy model for the QPOs in BH systems of different scales is proposed based on the magnetic reconnection of large-scale magnetic fields generated by the toroidal electric currents in the disk. The dynamical equations of accretion disk are resolved based on the interaction between the electric currents with the disk-corona system by using the iterative algorithm. The 190 Hz and 250 Hz single HFQPOs in BHBs XTE J1859+226 and XTE J1650$-$500 associated with SPL states are well fitted based on the disk-corona model with electric currents flowing in the inner disk. And the similar QPOs observed in ULX NGC 5408 X-1 and Seyfert 1 AGN RE J1034+396 and the corresponding X-ray spectra are also fitted. The spectrum of NGC 5408 X-1 is fitted with strong power-low component and steep power-law index suggesting that the QPO is similar to the HFQPOs in BHBs XTE J1859+226 and XTE J1650-500 and are probably associated with the same spectral state---SPL state. While the QPO in RE J1034+396 is analogous to the 67 Hz QPO in GRS 1915+105 which is associated with TD state.

Similarities of the QPOs in BH systems of different scales enable us to estimate some physical quantities of the massive BHs which have not been well constrained at present. As suggested by MR06, the frequencies of HFQPOs in BHBs showing the 3:2 frequency pairs are inverse proportional to the BH mass. Abramowicz et al. (\cite{abram07}) used the $1/M$ scaling to expect QPO frequencies for BHs of different scales and neutron stars. Similarly, we can use this relation to estimate the BH mass in ULX NGC 5408 X-1. Including the three BHBs showing the single HFQPO, we have the relationship, $m_{\textup{\scriptsize BH}}=1460\nu^{-1}_{\textup{\scriptsize QPO}}$, between the BH mass and the QPO frequency as shown in Fig. 10. The BH mass of NGC 5408 X-1 is then estimated about $7.3\times10^4M_\odot$ with the 0.02 Hz QPO as shown by the black dot in Fig. 10.

The quasi-periodic signals similar to QPOs were also discovered in the X-ray flux of Sgr A*. If these signals are indeed QPOs and are triggered by the magnetic reconnection described in this paper, then the lower limit of the BH spin can be constrained because the outer footpoints of the magnetic field lines cannot extend to the infinite distance. We estimate the lower limit of the BH spin of Sgr A* as 0.448 by fitting the 22.2 minutes signals discovered in the X-ray flare on 2004 August 31 (B¨¦langer et al. \cite{belanger}) using the mass $4.4\times10^6M_\odot$ (Genzel et al. \cite{genzel}), which is very close to the value $a_*\approx0.44\pm0.08$ estimated by Kato et al. (\cite{kato10}) using the QPO method in the context of disk-seismology.

It is noticed that the upper and lower kHz QPOs of accreting X-ray binaries are interpreted as the Alfven wave oscillations with different accreted material mass densities at a preferred radius near the star surface, and this model successfully explains the empirical relation between the upper and lower kHz QPO frequencies and the linear relation between the high and low QPO frequencies of BHs, neutron stars and white dwarfs (Zhang \cite{zhang04}; Zhang et al. \cite{zhang07}). The idea of interpreting QPOs with magnetic reconnection proposed in this paper may also apply to neutron stars and other astrophysical objects. For example, QPOs can be interpreted as the reconnection of magnetic field on the surface of a neutron star, then the model evolves to the beat frequency model (e.g., Miller et al. \cite{miller98}), providing a physical mechanism for QPO production.

Magnetic reconnection is being increasingly recognized as an important process in high energy objects, such as stellar X-ray flares, accretion disk corona, and magnetar flares. In this paper, we apply it to interpret the QPOs in BH systems of different scales based on a possible origin of the large-scale magnetic field in BH accretion disk. The magnetic fields generated in the inner disk by the electric currents are $\sim10^6$ and $\sim10^3$ Gauss for BHBs and AGNs respectively. Although the strength of these fields is much smaller than that of the tangled small-scale magnetic fields in the disk, the magnetic reconnection of the large-scale magnetic fields should have some effects on the heating of the corona, and these should be considered in the future work. Furthermore, the influence of the magnetic reconnection on X-ray spectra is another open question to be solved.

\begin{acknowledgements}
     We are very grateful to the anonymous referee for his (her) helpful comments on the manuscript. This work is supported by the NSFC (grants 11173011, 11143001, 11103003 and 11045004), the National Basic Research Program of China (2009CB824800) and the Fundamental Research Funds for the Central Universities (HUST: 2011TS159).
\end{acknowledgements}

\label{lastpage}


\begin{thebibliography}{}
  \bibitem[2007]{abram07} Abramowicz M. A., Klu$\acute{\textup{z}}$niak W., Bursa M., Hor$\acute{\textup{a}}$k J.,
  et al., 2007, RevMexAA (SC), 27, 8
  \bibitem[2004]{aschenbach} Aschenbach B., Grosso N., Porquet D., et al., 2004, A\&A, 417, 71
  \bibitem[2006]{belanger} B$\acute{\textup{e}}$langer G., Terrier R., de Jager A. C., et al., 2006, Journal of Physics: Conference Series, Volume 54, Proceedings of ``The Universe Under the Microscope - Astrophysics at High Angular Resolution", held 21-25 April 2008, in Bad Honnef, Germany. Editors: Rainer Schoedel, Andreas Eckart, Susanne Pfalzner and Eduardo Ros, pp. 420-426
  \bibitem[2010]{bian} Bian W.-H., \& Huang K., 2010, MNRAS, 401, 507
  \bibitem[1977]{bz77} Blandford R. D., \& Znajek R. L., 1977, MNRAS, 179, 433
  \bibitem[1982]{bp82} Blandford R. D., \& Payne D. G., 1982, MNRAS, 199, 883
  \bibitem[2004]{cropper} Cropper M., Soria R., Mushotzky R. F., et al., 2004, MNRAS, 349, 39
  \bibitem[2000]{cui} Cui W., Shrader C. R., Haswell C. A., et al., 2000, ApJ, 535, L123
  \bibitem[1999]{fender} Fender R. P., Garrington S. T., McKay D. J., et al., 1999, MNRAS, 304, 865
  \bibitem[2009]{gan09} Gan Z.-M., Wang D.-X., Lei W.-H., 2009, MNRAS, 394, 2310 (G09)
  \bibitem[2007]{gan07} Gan Z.-M., Wang D.-X., Li Y., 2007, MNRAS, 376, 1695
  \bibitem[2010]{genzel} Genzel R., Eisenhauer F., Gillessen S., 2010, Rev. Mod. Phys., 82, 3121
  \bibitem[2008]{gierlinski} Gierli$\acute{\textup{n}}$ski M., Middleton M., Ward M., 2008, Nature, 455, 369
  \bibitem[2003]{homan03} Homan J., Klein-Wolt M., Rossi S., et al., 2003, ApJ, 586, 1262
  \bibitem[2006]{homan06} Homan J., Wijnands R., Kong A., 2006, MNRAS, 366, 235
  \bibitem[2010]{huang} Huang C.-Y., Gan Z.-M., Wang J.-Z., Wang D.-X., 2010, MNRAS, 403, 1978
  \bibitem[2002]{karachentsev} Karachentsev I. D., Sharina M. E., Dolphin A. E., et al., 2002, A\&A, 385, 21
  \bibitem[2010]{kato10} Kato Y., Miyoshi M., Takahashi R., et al., 2010, MNRAS, 403, L71
  \bibitem[2002a]{li02a} Li L.-X., 2002a, ApJ, 567, 463
  \bibitem[2002b]{li02b} Li L.-X., 2002b, Phys. Rev. D, 65 084047
  \bibitem[1979]{linet} Linet B., 1979, J. Phys. A, 12, 839
  \bibitem[2002]{liubf} Liu B.-F., Mineshige S., Shibata K., 2002, ApJ, 572, L173
  \bibitem[2007]{liudm} Liu D.-M., Ye Y.-C., Wang D.-X., 2007, CTP, 47, 374
  \bibitem[1982]{macdonald} Macdonald D., \& Thorne K. S., 1982, MNRAS, 198, 345
  \bibitem[2010]{maitra} Maitra D., \& Miller J. M., 2010, ApJ, 718, 551
  \bibitem[2006]{mr06} McClintock J. E., \& Remillard R. A., 2006, in Lewin, van der Klis, eds, Compact Stellar X-ray Sources. Cambridge Univ. Press, Cambridge, p. 157 (MR06)
  \bibitem[2006]{mcclintock} McClintock J. E., Shafee R., Narayan R., 2006, ApJ, 652, 518
  \bibitem[2009]{middleton09} Middleton M., Done C., Ward M., et al., 2009, MNRAS, 394, 250
  \bibitem[2010]{middleton10} Middleton M., \& Done C., 2010, MNRAS, 403, 9
  \bibitem[1998]{miller98} Miller M. C., Lamb F. K., Psaltis D., 1998, ApJ, 508, 791
  \bibitem[2002]{miller02} Miller J. M., Fabian A. C., Wijnands R., et al., 2002, 570, L69
  \bibitem[2003]{miller} Miller J. M., Fabbiano G., Miller M. C., Fabian A. C., 2003, ApJ, 585, L37
  \bibitem[2004]{miniutti} Miniutti G., Fabian A. C., Miller J. M., 2004, MNRAS, 351, 466
  \bibitem[1997]{morgan} Morgan E. H., Remillard R. A., Greiner J., 1997, ApJ, 482, 993
  \bibitem[2004]{orosz} Orosz J. A., McClintock J. E., Remillard R. A., et al., 2004, ApJ, 616, 376
  \bibitem[2006]{remillard} Remillard R. A., \& McClintock J. E., 2006, ARA\&A, 44, 49
  \bibitem[2005]{roberts} Roberts T. P., Warwick R. S., Ward M. J., et al., 2005, MNRAS, 357, 1363
  \bibitem[1983]{Shapiro} Shapiro S. L., \& Teukolsky S. A., (1983). Black Holes, White Dwarfs and Neutron Stars, (John Wiley and Sons, Inc. New York). P.357
  \bibitem[2003]{strohmayer03} Strohmayer T. E., \& Mushotsky R. F., 2003, ApJ, 586, L61
  \bibitem[2009]{strohmayer09} Strohmayer T. E., \& Mushotsky R. F., 2009, ApJ, 703, 1386
  \bibitem[2007]{strohmayer07} Strohmayer T. E., Mushotzky R. F., Winter L., et al., 2007, ApJ, 660, 580
  \bibitem[2000]{wang} Wang D.-X., 2000, GRG, 32, 553
  \bibitem[2002]{wang02} Wang D. X., Xiao K., Lei W. H., 2002, MNRAS, 335, 655 (W02)
  \bibitem[2001]{wilms} Wilms J., Reynolds C. S., Begelman M. C., et al., 2001, MNRAS, 328, L27
  \bibitem[2008]{wu} Wu Q.-W., \& Gu M.-F., 2008, ApJ, 682, 212
  \bibitem[2000]{zdziarski} Zdziarski, A. A., 2000, in Highly Energetic Physical Processes, Procs. IAU Symposium $\sharp$195, eds. C. H. Martens, S. Tsuruta and M. A. Weber, ASP, 153-170
  \bibitem[2004]{zhang04} Zhang C.-M., 2004, A\&A, 423, 401
  \bibitem[2007]{zhang07} Zhang C.-M., Yin H.-X., Zhao Y.-H., 2007, PASP, 119,393
  \bibitem[2009]{zhao} Zhao C.-X., Wang D.-X., Gan Z.-M., 2009, MNRAS, 398, 1886 (Z09)
  \bibitem[2010]{zhou} Zhou X.-L., Zhang S.-N., Wang D.-X., et al., 2010, ApJ, 710, 16
  \bibitem[1978]{znajek} Znajek R. L., 1978, MNRAS, 182, 639
  \bibitem[2002]{zurita} Zurita C., Sanchez-Fernandez C., Casares J., et al., 2002, MNRAS, 334, 999

\end{thebibliography}
\end{document}